\documentclass[lettersize,journal]{IEEEtran}
\ifCLASSOPTIONcompsoc
\usepackage[nocompress]{cite}
\else
\usepackage{cite}
\fi

\ifCLASSINFOpdf
\else
\fi

\usepackage{amsmath,amsthm,amssymb,amsfonts,bm}
\newtheorem{theorem}{Theorem}
\newtheorem{observation}{Observation}
\newtheorem{lemma}{Lemma}
\newtheorem{proposition}{Proposition}

\newtheorem{corollary}{Corollary}
\newtheorem{defn}{Definition}
\usepackage{algorithmic}
\usepackage{float}
\usepackage{ragged2e}
\usepackage{graphicx}
\usepackage{epstopdf}
\usepackage{textcomp}
\usepackage{xcolor}
\usepackage{url}
\usepackage{enumitem}
\usepackage{url}
\setlist[itemize]{leftmargin=*}
\setlist[enumerate]{leftmargin=*}
\usepackage{multirow}
\newcommand{\tabincell}[2]{\begin{tabular}{@{}#1@{}}#2\end{tabular}} 
\usepackage{subfigure}
\usepackage[linesnumbered, ruled]{algorithm2e}
\begin{document}

\title{Information Disclosure under Competition in Sharing Systems}
\author{Ningning Ding, 
	Zhixuan Fang, 
	and~Jianwei~Huang,~\IEEEmembership{Fellow,~IEEE}
	\IEEEcompsocitemizethanks{\IEEEcompsocthanksitem Ningning Ding is with the Department of Electrical and Computer Engineering, Northwestern University, Evanston, IL 60208 USA.
		\IEEEcompsocthanksitem Zhixuan Fang is with  Institute for Interdisciplinary Information Sciences, Tsinghua University, Beijing, China, and Shanghai Qi Zhi Institute, Shanghai, China.
		\IEEEcompsocthanksitem Jianwei Huang is with the School of Science and Engineering, The Chinese University of Hong Kong, Shenzhen, China and the Shenzhen Institute of Artificial Intelligence and Robotics for Society (AIRS) (e-mail: jianweihuang@cuhk.edu.cn, corresponding author). } 
	\markboth{IEEE SYSTEM JOURNAL}
}

\maketitle

\begin{abstract}
Sharing systems have facilitated the redistribution of  underused resources  by providing convenient online marketplaces for individual sellers and buyers. However,  sellers in these systems may not   fully disclose the information of their shared commodities, due to strategic behaviors or privacy concerns. Sellers' strategic information disclosure  significantly affects buyers' user experiences and systems' reputation. This paper presents  the first analytical study on information disclosure and pricing  of competing sellers in  sharing systems. In particular, we propose a two-stage game framework to capture sellers' strategic behaviors and buyers' decisions. Although the  optimization problem is challenging due to sellers' non-convex and non-monotonic objectives, we  completely characterize the complex market equilibria by decomposing it into several tractable  subproblems.  We demonstrate  that full disclosure by all sellers or non-disclosure by all sellers will both lead to intense price competition. The former all-disclosure case is never an equilibrium even when all sellers have good commodity qualities and low privacy  costs, while the latter non-disclosure case can be an equilibrium under which all sellers get zero profit. We also reveal  several critical factors that affect sellers' information disclosure. Interestingly, sellers’ sharing capacity limitation and buyers' estimation biases encourage information disclosure as they mitigate sellers' competition.
\end{abstract}


\section{Introduction}
\label{sec:introduction}
\subsection{Background and Motivations}
\IEEEPARstart{T}{he} past few years have witnessed the fast-growing of sharing systems, such as network sharing (e.g.,   leftover mobile data trading \cite{Yu2020,2cm} and user-provided networks \cite{Khalili2015,bewifi}),  accommodation sharing  (e.g., Airbnb \cite{airbnb} and HouseTrip \cite{house}),  and labor sharing (e.g., TaskRabbit \cite{task}). 
In these systems, the sellers can profit from sharing their underused  resources (such as network resources,  houses, and personal time), and buyers can pay to access these resources through mobile phones \cite{Hamari2015}. A common  practice in sharing systems is that sellers display commodity information and announce the prices, and  buyers make choices accordingly \cite{Singh2018}. 
For example, in the network sharing systems (such as BeWiFi \cite{bewifi}), sellers with leftover network resources can upload descriptions (e.g.,  mobile network operators, network conditions, and locations) to demonstrate their commodity qualities and attract buyers, and buyers can purchase suitable resources. 
Due to  buyers' mobility, it is difficult for buyers to precisely  know sellers' commodity quality, and thus buyers rely heavily on sellers' disclosed information to make purchase decisions.

However, sellers may not be willing to show all their information for some reasons such as  strategic behaviors and privacy concerns. First, strategic sellers can manipulate  the  display effects   of commodities (which will affect buyers' perceptions) by controlling the degree of information disclosure  \cite{Balau2016}. Specifically, by disclosing information,  a seller  can display his competitive quality or distinguishing features to attract buyers when competing with other sellers. Meanwhile, the seller can also intentionally hide some unfavorable information from buyers, e.g., a host can conceal a photo of the living room which has a window facing a busy (hence potentially noisy) street. Second,  as   information can be widely spread, e.g., forwarded and shared through the mobile networks, social media,  and social networks, sellers may have privacy concerns when sharing personal properties to strangers  \cite{Liang2019}. In sharing systems, sellers  are often individuals instead of companies,  commodities are often personal idle assets/service rather than industrial products, and sellers still use the sharing resources by themselves occasionally.  Thus, information disclosure in sharing systems is always accompanied by  privacy disclosure, such as the locations in network sharing, home details in accommodation sharing,  and personal information in labor sharing, which explains sellers' incomplete disclosure behaviors \cite{Lutz2017}.  


Hence,  buyers usually have limited information about the qualities of commodities in sharing systems. They  have to make decisions based on their observation and estimation, i.e., observe what sellers show and estimate  what sellers do not show based on their  beliefs. 
Although some studies (e.g., \cite{Ma2017}) show that the public review is also an important source to know the qualities of commodities, 
some literature (e.g., \cite{Zervas2015,Fradkin2018})   reveals that reviews  on social networks  are always  over-positive or uninformative. Moreover, for the fresh sellers  in the system or  changeable commodities (e.g., unused network resource), the reviews are not helpful enough. 
Thus, as a critical approach for buyers evaluating the commodities, sellers' information disclosure levels directly and significantly affect buyers' user experiences and systems' overall reputation  \cite{Kim2015WhyPP}. We will focus on studying the impact of sellers' information disclosure in this paper, assuming that they have similar reputations (i.e., public perception from buyers)\footnote{We will consider  more complicated interactions between sellers'  disclosure strategies and buyers' reviews in a dynamic environment in future work.}. 

Understanding the  key factors of sellers' information disclosure  can  facilitate sharing systems' incentive mechanism design,  credibility evaluation, and  user loyalty improvement \cite{xie2017impacts}. However, few results have been  provided in  literature to elaborate the influencing factors of sellers' disclosure behaviors in sharing systems. 
A recent numerical study indicates that house quantity negatively moderates sellers' information disclosure intentions in a sharing system \cite{Liang2019}. Nevertheless,  the impact of such capacities on sellers' information disclosure strategies has not been analytically studied. Moreover, sharing systems  have recently released some new regulations on  quality improvement and capacity limitation. For example, Airbnb has launched the Airbnb Select program to improve property qualities \cite{tim} and has restricted hosts from renting out multiple properties in some areas  \cite{qtimmurphy.org}.  It is not clear yet what impact these regulations will have on sellers' information disclosure. Therefore, this paper will also study the impact of some key factors on sellers' optimal information disclosure decisions.

\vspace{-3mm}
\subsection{Model and Problem Formulation}
\vspace{-1mm}
We consider a general  sharing system where sellers and buyers are individuals, which applies to a broad range of sharing scenarios.
In such a sharing system, each seller decides the disclosed information about his shared commodities and sets a price. Each buyer compares the displayed information and prices of the sellers, and makes a choice among the sellers to maximize his expected payoff. 
In this paper, we would like to answer the following key questions: 
\begin{itemize}
	\item 
	\emph{How do  privacy-sensitive  sellers  disclose  information under competition in sharing systems?}
	\item 
	\emph{How is sellers' optimal information disclosure  affected by the factors such as sellers' commodity qualities, privacy costs,  capacity limitation, and buyers' estimation biases?}
\end{itemize}
\vspace{-3mm}
\subsection{Contributions}
\vspace{-1mm}
To the best of our knowledge, this is the first work that conducts an in-depth theoretical   analysis of sellers' information disclosure and price competition in sharing systems. 
We summarize our key contributions as follows:
\begin{itemize}
	\item We propose a game-theoretic model to characterize the interactions between  sellers and buyers under incomplete  information disclosure in sharing systems. The model captures both  buyers' decision processes (observation and estimation) and sellers'  strategic behaviors (trade-offs between the commodity display effects  and privacy concerns). 
	\item We solve a challenging non-convex optimization problem involving piece-wise functions derived from the two-stage game, and give a complete characterization on the complex structure of equilibria under different conditions. 
	\item  We reveal the relationship between information disclosure and price competition. The results show that  full information disclosure by all sellers is never an equilibrium due to fierce price competition, even if all sellers have good commodity qualities and low privacy costs. Meanwhile, non-disclosure by all sellers leads to intense price competition and makes them get zero profit, which happens when each  seller has a poor commodity quality or an unaffordable privacy cost. 
	\item We characterize the key factors of sellers' disclosure strategies.  Limited sharing capacities and the existence of buyers'  estimation biases lead to less intense  competition among sellers as well as more information disclosure in systems; high commodity qualities and low privacy costs also encourage sellers' information disclosure.
\end{itemize}

\vspace{-2mm}
\subsection{Related Work}
\vspace{-1mm}
\label{literature}
Our paper is mostly related to the following  two distinct strands of literature.
\subsubsection{Sharing Systems}
Studies in sharing systems have sprung up in recent years. 
Those related studies on the economics aspects of sharing systems have focused on several key issues such as incentives (e.g., \cite{nie2020multi,sedghani2021incentive}), pricing (e.g., \cite{Zheng2020,Lu2020}), throughput (e.g., \cite{Courcoubetis2019,Lin2019}), subsidization (e.g., \cite{Fang2017,fang2020loyalty}), popularity prediction (e.g., \cite{Ouyang2019}), and matching (e.g., \cite{Peng2018,Fraiberger2015}). Some of them also considered   two-stage Stackelberg games under incomplete information  but focused on different research problems from ours (e.g., \cite{nie2020multi,sedghani2021incentive}). In particular, Vivek \emph{et al.}  \cite{Singh2018} and Xu \emph{et al.} \cite{xu2021impact} numerically  investigated how sellers' strategic signals  impact buyers' renting decisions on Airbnb though empirical data. However, there does not exist a comprehensive analytical study on this issue yet. We provide a complementary	    view to fill this gap in the literature.  To the best of our knowledge, our paper is the first attempt that analytically studies sellers' information disclosure in sharing systems, which provides a clear understanding regarding the observations in the related numerical studies \cite{Singh2018,xu2021impact}.

\subsubsection{Information Disclosure under Competition}

Although there is a rich body of literature on sellers' information disclosure, 
the research about sellers' information disclosure under competition has only recently emerged  with several modeling limitations. Some studies  assumed that the buyers are homogeneous (e.g., \cite{D3,D4}) or disclosure is costless (e.g., \cite{D5,D6}). Only a few of them  considered the uncertainty of buyers' decisions,  albeit in different ways from this work (e.g., \cite{D1,D2,D5,D6}). A widely adopted setting is the two-seller setting which is less general than the models with  multiple sellers (e.g., \cite{D1,D2,D3,D4,D6}). 
Many works used  static models, but multi-stage dynamic models are  closer to natural processes of interactions among users (e.g., \cite{D4,D5}).  Moreover, there is a lack of analytical research about the impact of sellers' capacity limitation on their  disclosure strategies. Although some models used the signaling game which can also  capture the interaction process (e.g., \cite{D2}), those work did not consider many other aspects of our problem setting (e.g., capacity limitations or multiple sellers).  As an attempt to generalize these diverse models, we incorporate    all these factors in our model, which is summarized in Table \ref{tablit}. As we shall see in next section, our model also  captures the unique features of sharing systems, such as users' interaction processes and privacy concerns.
\begin{table}[tbp]
	\caption{Comparison of Models in the Literature}  
	\vspace{-5mm}
	\begin{center}
		\begin{tabular}{|c|p{0.3cm}cccc|}
			\hline
			Ref.& Cost&\tabincell{c}{Buyer\\Uncertainty}& \tabincell{c}{Capacity\\Limit}&\tabincell{c}{Multiple\\Sellers} &\tabincell{c}{Multi-\\Stage\\ Model}\\
			\hline
			\cite{BOARD2009}  &$\,\times$& $\times$&$\times$&$\times$ &$\times$\\
			\hline
			\cite{Cheong2004}  &$\,\surd$  &$\times$ &$\times$&$\surd$ &$\times$\\
			\hline
			\cite{Janssen2016} & $\,\times$& $\surd$ &$\times$&$\times$ &$\surd$\\
			\hline
			\cite{D6} & $\,\times$& $\surd$ &$\times$&$\times$ &$\times$\\
			\hline
			\cite{Stivers2004} & $\,\times$& $\,\surd$ &$\times$&$\surd$ &$\surd$\\
			\hline
			\tabincell{c}{\cite{D1,D2}} &$\,\surd$  & $\surd$  &$\times$& $\times$&$\times$\\
			\hline
			\cite{D3} &$\,\surd$  & $\times$  &$\times$& $\times$&$\times$\\
			\hline
			\cite{D4} &$\,\surd$  & $\times$  &$\times$& $\times$&$\surd$ \\
			\hline
			\cite{D5} &$\,\times$  & $\surd$   &$\times$& $\surd$ &$\surd$ \\
			\hline
			\textbf{This paper} & $\,\bm{\surd}$ & $\bm{\surd}$& $\bm{\surd}$&$\bm{\surd}$ &$\bm{\surd}$\\
			\hline
		\end{tabular}
		\label{tablit} 
	\end{center}
	\vspace{-8mm}
\end{table}

Furthermore, there has been a heated debate about the disclosure strategies of sellers in competition. Some works (e.g., \cite{D3,D4}) showed  that competing sellers will fully disclose information if there is no cost.  However, empirical studies (e.g., \cite{Dedman2009}) indicated that  sellers may withhold information  if they perceive strong competition.  A number of analytical studies also pointed out that competitive pressures between  sellers can undermine the full disclosure result, and only high-quality  sellers  disclose information (e.g., \cite{D6}). 
Moreover, some studies took a different view that no disclosure by any seller may  be the unique outcome even if disclosure cost vanishes (e.g., \cite{Cheong2004}). Our paper provides more general  analysis that captures these results as special cases and completely characterizes the  different equilibria under various conditions. 

The rest of the paper is organized as follows. We  describe the system model in Section \ref{model}. We first consider a basic setting with two sellers  and one buyer   in Section \ref{pre}. In Section \ref{multiple}, we analyze the more general case with multiple sellers and buyers, taking sellers' capacities into consideration.  
Finally, we  conclude in Section  \ref{conclusion}.

\vspace{-3mm}
\section{System Model}
\vspace{-1.5mm}
\label{model}

We consider a sharing system   where sellers rent out the same type of idle resources\footnote{As a concrete example, a buyer visiting a city at a specific time  wants to book an apartment in a specific area in an accommodation sharing system. In this case, the same type of resources mean that the apartments of competing sellers have the same properties in terms of available time and location. Hence, these apartments are substitutes for each other.} for monetary reward and buyers pay to access the resources according to their needs.  In Sections \ref{model} and \ref{pre}, we first consider the case where the system consists of two sellers  (seller 1 and seller 2) and one buyer. The \emph{true quality} of seller $i$'s  commodities (service) is denoted by $Q_i$ ($i \in \{1,2\}$), which is exogenously determined. Without loss of generality, we assume that seller 1 provides higher quality commodities than seller 2, i.e., $Q_1\ge Q_2\ge 0$. 

The two-seller setting is widely adopted by  literature (e.g., \cite{D3,Janssen2016,D6}). The one-buyer setting is  equivalent to the case where multiple heterogeneous buyers come to the system one at a time when sellers  have unlimited capacities. 	
We will further study the general case with $N$ sellers and $K$ buyers in both unlimited and limited capacity scenarios in Section \ref{multiple}.

\vspace{-3mm}
\subsection{Strategies of Sellers and Buyers}
\vspace{-1mm}
\subsubsection{Sellers' Information Disclosure and Pricing}
As a common policy of sharing systems, each seller needs to decide how much \emph{private information} (e.g., workers' educational levels, rooms' decoration styles) of his commodities  to disclose (to buyers) and what  commodity price to set.

We let $\alpha_i \in [0,1]$ denote the seller $i$'s \emph{disclosure level}, i.e., the fraction of seller $i$'s disclosed private information. The case of $\alpha_i=1$ means that seller $i$  discloses all the private information and shows  the true quality to buyers\footnote{We assume that the disclosed information is real and thus reflects the true quality, because sharing systems usually impose severe penalties on sellers who disclose fake information  \cite{policy,term}.}. The case of $\alpha_i \in (0,1)$ means that seller $i$ voluntarily  discloses part of his information.   The case of $\alpha_i=0$ means that seller $i$ discloses  minimum  information  according to the regulation of the system. However, for the convenience of presentation, we describe $\alpha_i=0$ as \emph{no  disclosure} throughout this paper. 
When making decisions on information disclosure, each seller has a trade-off between the commodity display effect for attracting buyers and the  privacy loss brought by such  disclosure. 
Thus, sellers may adopt incomplete information disclosure strategies, e.g., a worker on TaskRabbit may just sketch his educational experience to protect his privacy and conceal the unfavorable information. 
As the first attempt to understand such a tradeoff, we assume that buyers can infer the  sellers'  information disclosure levels  through observations\footnote{A seller’s disclosure level can be observed consistently by all buyers. For example, suppose that there are five pieces of information about  network sharing  that sellers can choose to disclose, including  the name of the network operator, the total network bandwidth, the total network coverage, the average signal quality, and the average network delay. If a seller only posts the information about coverage and delay on his website, then we can treat the seller's  information disclosure level as $\alpha=0.4$.}.

Besides choosing the information disclosure level, each seller $i$ also needs to choose  a  price $p_i$ for his commodities. Setting a high price will decrease the competitiveness of  the seller, while setting a low price may make the seller unprofitable.

\subsubsection{Buyers' Choice}
The buyer compares the expected qualities of the commodities provided by the two  sellers as well as the corresponding prices, and then chooses the  seller that gives her maximum payoff (defined in Section \ref{buyerpayoff}). We assume that  the  true qualities of sellers satisfy the buyer's minimum requirement and the buyer only buys one  commodity  from the chosen seller. 

\vspace{-3mm}
\subsection{Buyers' Payoff}
\vspace{-1mm}
\label{buyerpayoff}
When the buyer  chooses seller $i\in \{1,2\}$, the buyer's payoff  is the difference between her expected commodity quality provided by the seller and her payment to the seller.  

The expected commodity quality depends on seller $i$'s information disclosure strategy:
\begin{itemize}
	\item If seller $i$ fully discloses his information, then the buyer knows that she will experience  the true quality $Q_i$\footnote{We assume that  buyers' different preferences (e.g., on decoration style of houses) do not affect their perception on sellers' qualities.}.
	\item If  seller $i$ discloses no information, then the buyer perceives the quality only by her estimation $Q_0+\epsilon$. Here, constant $Q_0$ is buyers' average prior belief on sellers' qualities  when the sellers do not disclose information. Random variable $\epsilon$ indicates the bias of the buyer's estimation. We assume that  $\epsilon$ obeys a uniform distribution, i.e., $\epsilon \sim U(-\epsilon_0,\epsilon_0)$, where $\epsilon_0$ is the maximum deviation from the average prior belief on quality and is assumed to satisfy $\epsilon_0 \le Q_0/3$.\footnote{This assumption is  not restrictive. First, it is reasonable that buyers in the same sharing system have similar  estimation on unknown commodity qualities. Second, literature shows that there is a pull-to-center effect of people's estimation, i.e., the estimated values of the majority  (71\% \cite{Christenfeld1995}, 76.2\% \cite{TEIGEN1983}) are the middle values.}  
	\item If  seller  $i$ partially discloses his information, the buyer's expected quality is modeled as $\alpha_i Q_i + (1-\alpha_i) (Q_0 + \epsilon)$. For the revealed part (fraction $\alpha_i$), the buyer observes quality $Q_i$; while for the unobservable part (fraction $1-\alpha_i$), the buyer estimates the quality with $(Q_0 + \epsilon)$. 
\end{itemize}  

Hence, when the  buyer   chooses seller $i$, the buyer's payoff is as follows:
\begin{equation}
U_i = \alpha_i Q_i + (1-\alpha_i) (Q_0 + \epsilon)  - p_i,
\; i\in \{1,2\}.
\label{equ:1}
\end{equation}
One can also understand the $\alpha_i$ as the weight/probability of the buyer believing in the sellers' displayed quality. Then, Equation \eqref{equ:1} means that more disclosed information (i.e., a larger $\alpha_i$) will increase the buyer's belief in the commodity quality and reduce buyers' own estimation. Such a linear expected commodity quality in \eqref{equ:1} is also used by recent studies \cite{Li2017,Ren2013}, showing that individuals are likely to estimate the unknown details based on their own knowledge/experiences rather than the advertised part. 
Such a behavior of distrust is natural in sharing systems, since the commodities are from various and  strategic individual sellers \cite{Liang2017,Tussyadiah2014,So2018}.\footnote{The  payoff function \eqref{equ:1} can be generalized to the more practical case  where sellers' displayed qualities deviate from the true qualities. 
	Assume that seller $i$'s displayed quality is ($Q_i+\delta$) instead of $Q_i$, where $\delta$ is the difference between the displayed quality and the true quality. Buyers do not know the value of $\delta$ and have to maximize the expected payoff. Their expected payoff is still \eqref{equ:1} if $\delta$ is a zero-mean variable, i.e., $E[\delta] = 0$.}  
Note that the buyer  has a reservation payoff of zero, i.e., if all sellers offer the buyer negative payoffs, she will exit the market.   
\subsection{Sellers' Profits}
\label{host}
Each seller's expected profit  is the difference between his expected income charged from the buyer and his privacy cost.

Specifically, a seller's expected income is the product of the price and the probability of winning  the buyer. When facing a buyer with an  estimation bias of $\epsilon$, seller $i$'s winning  probability is jointly determined by the strategies of all sellers, and is denoted by  $Pr_i(\boldsymbol{\alpha}, \boldsymbol{p},\epsilon) \in [0,1]$, where $\boldsymbol{\alpha}=(\alpha_1,\alpha_2)$ are sellers' disclosure levels and $ \boldsymbol{p}=(p_1,p_2)$ are their prices. 
\begin{itemize}
	\item If seller $i$ gives the buyer  a higher payoff, then $Pr_i(\boldsymbol{\alpha}, \boldsymbol{p},\epsilon)=1$, i.e., seller $i$ wins the buyer.
	\item If seller $i$ offers the buyer a lower  payoff, then $Pr_i(\boldsymbol{\alpha}, \boldsymbol{p},\epsilon)=0$, i.e.,  seller $i$ loses the buyer.
	\item If two sellers give the same  payoff,  then $Pr_i(\boldsymbol{\alpha}, \boldsymbol{p},\epsilon)=0.5$, i.e., tie breaks randomly.
\end{itemize}   

The disclosure of information induces privacy/information costs for sellers  \cite{Verrecchia1983}. 
We assume that seller $i$'s  privacy cost  
is proportional to the private information disclosure level and is modeled as $c_i \cdot \alpha_i$, where  $c_i$ measures how seller $i$ values his privacy.  Such a linear privacy cost model has been widely used in literature (e.g., \cite{Ghosh2015,Ghosh2014}).\footnote{Information disclosure happens before the buyer purchasing the commodity, so sellers' information costs occur even if their commodities are not eventually sold. 
} 

Hence, when facing a buyer with an  estimation bias of $\epsilon$, seller $i$'s profit is summarized as follows\footnote{Our model can be	applied to the case where sellers’ profits include an additional homogeneous	service cost. Such a service cost will not affect the equilibrium results in this paper. Thus,  we normalize it to zero.}:
\begin{equation}
W_i(\boldsymbol{\alpha}, \boldsymbol{p},\epsilon)  = p_i \cdot Pr_i(\boldsymbol{\alpha}, \boldsymbol{p},\epsilon)  - c_i \cdot \alpha_i, \; i\in \{1,2\}.
\label{equ:2}
\end{equation}

The buyer's estimation bias $\epsilon$ is a random variable that is unknown to sellers, so  seller $i$ has to maximize his expected profit:
\begin{equation}
E[W_i(\boldsymbol{\alpha}, \boldsymbol{p})] = E_{\epsilon}[p_i \cdot Pr_i(\boldsymbol{\alpha}, \boldsymbol{p},\epsilon)  - c_i \cdot \alpha_i],\; i\in \{1,2\}. 
\label{equ:15}
\end{equation}
Sellers are involved in game theoretical interactions, as every seller's profit depends on  all sellers'  decisions. 
\vspace{-1.8mm}
\subsection{The Two-Stage Game}
\vspace{-0.7mm}
\label{threestage}
We model the strategic interactions  between sellers and buyers as a two-stage Stackelberg game. 
In Stage I, as the leaders, sellers simultaneously determine their information disclosure levels $\boldsymbol{\alpha}$  and announce the  prices $\boldsymbol{p}$ to the buyers.  In Stage II, as the follower, each buyer chooses one seller.  Such a model is applicable to both Sections \ref{pre} and \ref{multiple}. 

\begin{figure}[tbp]	
	\centering  	
	\includegraphics[width=0.65\linewidth]{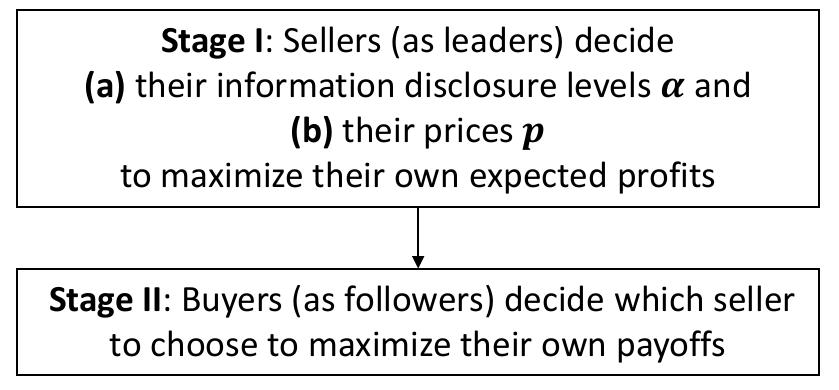} 
	\vspace{-4mm}
	\caption{The two-stage game.}  
	\label{fig:1}   
	\vspace{-6mm}
\end{figure}

This two-stage game  captures users' practical sequential interaction processes in sharing systems (Fig.~\ref{fig:1}). Specifically, sellers first show buyers their commodity information through photos and descriptions, which reflect their commodity qualities. They also announce the prices for their commodities.  Hence, we model the sellers as the leaders in Stage I. 
After observing sellers' information and prices, each buyer will pay for a commodity from the seller who gives her the highest payoff. Thus, buyers' decisions  are modeled in Stage II. Mathematically,  Stage I is a non-cooperative game among sellers, as each seller’s profit depends on both his strategies and also other   sellers’ strategies. In Stage II, as we consider buyers coming to the system one by one, so each buyer makes her own decision independent of other buyers.

We analyze the games between  sellers as complete information games, i.e., the privacy costs and commodity  qualities are known by all sellers (but not by the buyers)\footnote{Consider the example for accommodation sharing on Airbnb. Suppose that a buyer wants to compare sellers in a certain area. 
	In reality, sellers usually   have some knowledge about their neighbors' house (e.g., neighbors' house qualities, attitude/cost for information privacy) \cite{zhou2008preserving,hampton2007neighborhoods}. 
	The main reason is that sellers often have repeated interactions in the market, and as a result it is possible for them to learn each other's information over time, especially when the commodities are relative stable over time (such as the same houses). However, as the buyers are often one-time visitors to these areas, they do not know the seller's private information.}. We use  standard backward induction to analyze the dynamic game. 

Note that the analysis of Stage I is mathematically equivalent to sequentially optimizing the two variables in the backward induction. Specifically,  sellers first  simultaneously determine their information disclosure levels $\boldsymbol{\alpha}$  (a game of information disclosure) in Stage I-a, and then simultaneously announce the  prices $\boldsymbol{p}$ to the buyers in Stage I-b (a game of pricing).  

\begin{table}[tbp]
	\caption{Key Notations}  
	\vspace{-5mm}
	\begin{center}
		\begin{tabular}{|c|c|}
			\hline
			$Q_i$ &  True quality of seller $i$'s commodity \\
			\hline
			$c_i$ &  Marginal privacy/information cost of seller $i$  \\
			\hline
			$\alpha_i$ &  Information disclosure level of seller $i$ \\
			\hline
			$p_i$ &  Commodity price of seller $i$  \\
			\hline
			$Q_0$ &   Buyers' average prior belief on sellers' qualities\\
			\hline
			$\epsilon$ &   Variance/bias of buyers' estimation\\
			\hline
			$\epsilon_{0}$ & Maximum deviation of buyers' estimation from   average belief\\
			\hline
		\end{tabular}
		\label{table} 
		\vspace{-6.7mm}
	\end{center}
\end{table}
For the ease of reading, we list the key notations in Table \ref{table}. In the following, we will first consider a basic setting where there are two sellers and one buyer. After that, we will further study the general case with many sellers and buyers.

\section{Basic Setting:  Two Sellers and One Buyer }
\vspace{-1mm}
\label{pre}
In this section, we consider the basic setting with two sellers and one buyer in the system. 
We will use the backward induction to analyze the two-stage game, deriving the buyer's decision, sellers' prices, and sellers'  information disclosure levels at equilibrium, respectively. After that, we  study the  factors that influence sellers' information disclosure.

\vspace{-2.5mm}
\subsection{Buyer's Decision in Stage II}
\vspace{-1mm}
\label{customer}
In this subsection, we analyze the buyer's decision by comparing the payoffs of choosing different sellers. Specifically, the buyer will choose the seller $i^*$ who offers her a higher expected payoff in Stage II.

Recall that the buyer's payoff is defined in (\ref{equ:1}). Therefore, the buyer's optimal choice $i^*$ is:
\begin{equation}
i^*(\boldsymbol{\alpha},\boldsymbol{p}) = \mathop{\arg\max}_{i\in \{1,2\}} \ \ \alpha_i Q_i + (1-\alpha_i) (Q_0 + \epsilon)  - p_i.
\end{equation}
If the two sellers give the buyer the same payoff,  the buyer will select each seller with an equal probability of 0.5. 

For the convenience of presentation, we first introduce the following definition:
\begin{equation}
\psi(\alpha_i,\alpha_j) \triangleq \alpha_i (Q_i  - Q_0) - \alpha_j (Q_j-Q_0) , \; i,j \in \{1,2\}, \; i\neq j.
\label{advantage}
\end{equation}
The term $\psi(\alpha_i,\alpha_j)$ is the difference of the buyer's expected qualities between choosing seller $i$ and seller $j$ under no estimation bias ($\epsilon=0$). When $\psi(\alpha_i,\alpha_j)>0$,   seller $i$  gives the buyer a higher expected quality than seller $j$. 

Then, the probability of a seller being selected by the buyer is characterized by the following Lemma \ref{thm1}.
\begin{lemma}
	\label{thm1}
	Given a uniformly random estimation bias $\epsilon \sim U(-\epsilon_0,\epsilon_0)$\footnote{The analysis is similar if we consider a different distribution of the buyer's quality estimation bias $\epsilon$. The only difference is that when sellers calculated their expected profit, they need to take the expectation over a different distribution of $\epsilon$. The key insights will remain the same.}, the probability of seller $i$ being selected by the buyer is ($ i,j \in \{1,2\}, i \neq j$):
	\begin{equation}
	\begin{split}
	Pr_i(\boldsymbol{\alpha}, \boldsymbol{p},\epsilon) 
	= \left\{ 
	\begin{array}{rcl}
	0,      \; \rm{if }\; $$(\alpha_i - \alpha_j) \epsilon     >      \psi(\alpha_i,\alpha_j) -(p_i - p_j);$$\\
	\frac{1}{2},     \; \rm{if }\; $$(\alpha_i - \alpha_j) \epsilon     =      \psi(\alpha_i,\alpha_j) -(p_i - p_j);$$\\
	1,     \; \rm{if }\; $$(\alpha_i - \alpha_j) \epsilon     <      \psi(\alpha_i,\alpha_j) -(p_i - p_j).$$\\
	\end{array} \right.
	\end{split}
	\label{equ:p}
	\end{equation}
\end{lemma}
Proof of Lemma \ref{thm1} is given in Appendix I. Lemma \ref{thm1} shows that the buyer's decision  heavily depends on the  difference between two sellers' information disclosure levels ($\alpha_i-\alpha_j$) as well as the buyer's estimation bias $\epsilon$.
\begin{itemize}
	\item When seller $i$ discloses less information than seller $j$ ($\alpha_i<\alpha_j$), a larger $\epsilon$ helps increase seller $i$'s winning probability. Estimation bias weakens the effectiveness  of disclosed information. 
	\item When two sellers disclose the same amount of information ($\alpha_i=\alpha_j$), the value of $\epsilon$ does not affect the buyer's decision. The buyer's decision is only affected by sellers' true qualities and prices, as she has the same uncertainty regarding the two seller's true qualities.
\end{itemize}

Moreover, the value of $\psi(\alpha_i,\alpha_j) -(p_i - p_j)$ is the difference between two sellers' expected quality gap and price gap. Seller $i$'s winning probability  increases with the expected quality gap  and decreases with the price gap. Thus, whether a seller can win the buyer largely depends on two sellers'  disclosure strategies in Stage I-a and pricing strategies in Stage I-b .

\vspace{-2.5mm}
\subsection{Sellers' Pricing Competition in Stage I-b }
\vspace{-1mm}
\label{price} 
In this subsection, we analyze sellers' pricing strategies in Stage I-b. Given the information disclosure levels $ \boldsymbol{\alpha} =(\alpha_1,\alpha_2)$ in Stage I-a,  sellers need to decide their  prices ($p_1^*,p_2^*$)  under the uncertainty of $\epsilon$, considering the buyer's decision in Stage II. 

Recall that a seller's expected profit is defined in (\ref{equ:15}). Since the disclosure level $\alpha_i$ is fixed in Stage I-b , seller $i$ only needs to choose a non-negative price to maximize his expected profit. Formally, the Sellers' Pricing Game and the corresponding equilibrium are defined as follows.
\begin{defn}[Stage I-b : Sellers' Pricing Game]\quad\\
	\vspace{-4.5mm}
	
	\begin{itemize}
		\item Players: seller 1 and seller 2.
		\item Strategy space: each seller $i \in \{1,2\}$ chooses his price $p_i\in  [0,\infty)$. 
		\item Payoff function: each seller $i \in \{1,2\}$  maximizes his expected profit $E[W_i(p_i;p_j,\boldsymbol{\alpha})]$, where $j \in \{1,2\}, j\neq i$.
	\end{itemize}
	\label{game1}
\end{defn}
\begin{defn}[Pricing Equilibrium in Stage I-b ]
	The equilibrium of the Sellers' Pricing Game is a profile ($p_1^*,p_2^*$) such that for each seller $i\in \{1,2\}$,
	\vspace{-1mm}
	\begin{equation*}
	E[W_i(p_i^*,p_j^*;\boldsymbol{\alpha})] \ge E[W_i(p_i,p_j^*;\boldsymbol{\alpha})], \quad \forall p_i \in [0,\infty).
	\end{equation*}
	
\end{defn}

In other words, no seller wants to unilaterally change his pricing decision at an equilibrium. Next, we give the pricing strategies of two sellers at the equilibrium  ($p_1^*,p_2^*$), which are the functions of the information disclosure levels ($\alpha_1, \alpha_2$).


\begin{lemma}
	\label{thm2}
	Given the information disclosure levels $\boldsymbol{\alpha} =( \alpha_1, \alpha_2)$, the equilibrium of the Sellers' Pricing Game (Stage I-b ) is:
	\begin{enumerate}
		\renewcommand{\labelenumi}{i)}
		\item If $\alpha_1 \neq \alpha_2$,
		\begin{equation*}
		\begin{split}
		&(p_1^\ast(\boldsymbol{\alpha}), p_2^\ast(\boldsymbol{\alpha})) =\\ 
		&\hspace{3.9mm}\begin{cases}
		(0, -\psi(\alpha_1,\alpha_2) - \xi),   \hspace{11.5mm} \rm{if }\; $$\psi(\alpha_1,\alpha_2) \le -3\xi,$$\\
		( \frac{3\xi+\psi(\alpha_1,\alpha_2)}{3} ,  \frac{3\xi-\psi(\alpha_1,\alpha_2)}{3} ),    \hspace{1.4mm} \rm{if }\; $$- 3\xi <\psi(\alpha_1,\alpha_2) < 3\xi,$$\\
		(\psi(\alpha_1,\alpha_2) - \xi ,  0) ,     \hspace{13.6mm}   \rm{if }\; $$\psi(\alpha_1,\alpha_2) \ge 3\xi,$$\\
		\end{cases}
		\end{split}
		\label{equ:3}
		\end{equation*}
		where $\xi \triangleq \epsilon_0|\alpha_1 - \alpha_2|$.
		\renewcommand{\labelenumi}{ii)}
		\item If $\alpha_1 = \alpha_2> 0$, there is no pure strategy equilibrium.
		\renewcommand{\labelenumi}{iii)}
		\item If $\alpha_1 = \alpha_2 = 0$,
		$p_1^*=p_2^*=0 $.
	\end{enumerate}
\end{lemma}
\begin{figure}[tbp]	
	\centering  	
	\includegraphics[width=0.5\linewidth]{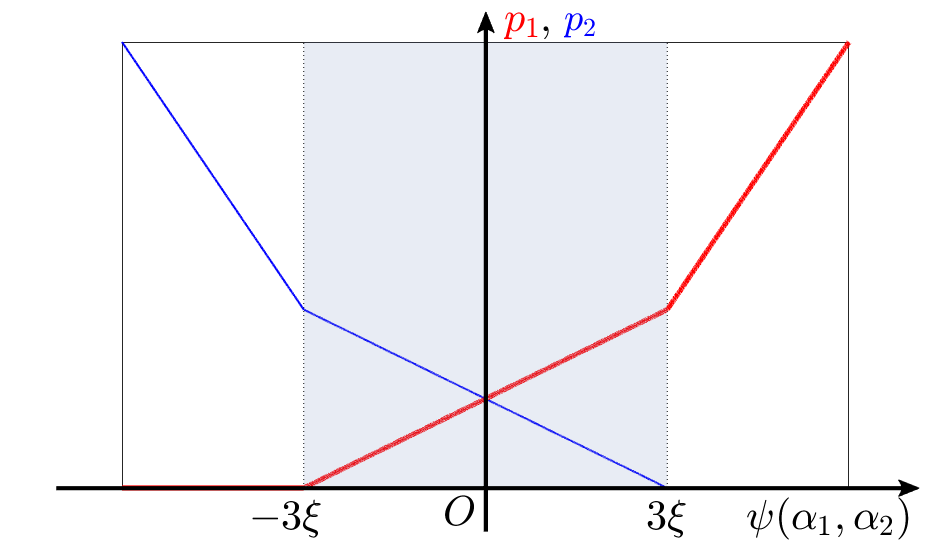} 
	\vspace{-3mm}
	\caption{Sellers' prices at the equilibrium $(p_1^\ast(\boldsymbol{\alpha}), p_2^\ast(\boldsymbol{\alpha}))$ when $\alpha_1 \neq \alpha_2$.}  
	\label{fig:price}   
	\vspace{-6mm}
\end{figure} 
Proof of Lemma \ref{thm2} is given in Appendix II. 
Here we list some interesting observations.
\begin{itemize}
	\item Fig.~\ref{fig:price} shows sellers' optimal prices when they set different disclosure levels ($\alpha_1 \neq \alpha_2$). If the expected quality gap is large ($|\psi(\alpha_1,\alpha_2)|  \ge 3\xi$), the seller who offers a higher expected quality will set a positive price, and his competitor (the other seller) will set a zero price. If the expected quality gap is small ($|\psi(\alpha_1,\alpha_2) |< 3\xi$), they will both set positive prices.  We shall see in  next subsection that the seller who sets a positive price has a positive expected profit; the seller who sets a zero price never discloses information and gets a zero profit.
	\item  Both sellers disclosing same amount of information ($\alpha_1 = \alpha_2= \alpha>0$) is never  an equilibrium. If their disclosure levels are the same in Stage I-a,  the difference of buyer's payoffs between choosing two sellers is $U_1-U_2 = \alpha(Q_1-Q_2) -(p_1-p_2)$. Since (without loss of generality we have assumed)  $Q_1\ge Q_2$, the ensuing price competition in Stage I-b  will lead seller 2 to  a zero price and  a negative profit $-c_2\alpha_2$, which is worse than no disclosure. 
	\item If no seller discloses information ($\alpha_1 = \alpha_2 = 0$), then $U_1-U_2 = p_2-p_1$. There will be an intense price competition where both sellers set a zero price at the equilibrium.
\end{itemize}

Note that sellers' pricing decisions depend on information disclosure levels, which will be analyzed in  next subsection.

\vspace{-1.5mm}
\subsection{Sellers' Information Disclosure in Stage I-a} 
\vspace{-0.5mm}
\label{information}
In this subsection, we study the sellers'  information disclosure strategies in Stage I-a. Considering the pricing strategies in Stage I-b  and the buyer's decision in Stage II, the sellers need to decide the information disclosure levels   ($\alpha_1,\alpha_2$) in Stage I-a to maximize their expected profits. 

Formally, the Sellers' Information Disclosure Game and the corresponding equilibrium are defined as follows.
\begin{defn}[Stage I: Sellers' Information Disclosure Game]\quad \\
	\vspace{-9mm}
	\begin{itemize}
		\item Players: seller 1 and seller 2.
		\item Strategy space: each seller $i\in \{1,2\}$ chooses his information disclosure level $\alpha_i \in  [0,1]$.
		\item Payoff function: each seller $i\in \{1,2\}$ maximizes his expected profit $E[W_i(\alpha_i;\alpha_j)]$,  where $j \in \{1,2\}, j\neq i$.
	\end{itemize}
	\label{game2}
\end{defn}
\begin{defn}[Disclosure Equilibrium in Stage I]
	The equilibrium of the Sellers' Information Disclosure Game is a profile ($\alpha_1^*,\alpha_2^*$) such that for each seller $i\in \{1,2\}$,
	\begin{equation*}
	E[W_i(\alpha_i^*,\alpha_j^*)] \ge E[W_i(\alpha_i,\alpha_j^*)], \quad \forall \alpha_i \in [0,1].
	\end{equation*}
\end{defn}

To obtain  sellers' expected profit function $E[W_i(\alpha_i;\alpha_j)]$ in Stage I-a, we substitute the results of Stage II (Lemma \ref{thm1}) and Stage I-b  (Lemma \ref{thm2}) into \eqref{equ:15}, as well as calculating the expectation over random bias $\epsilon$. It turns out to be a challenging   optimization problem in Stage I-a  due to sellers' complex non-convex objective functions (as shown in Equations (14) and (15) in Appendix III).   The challenge is  induced by i) the discontinuity of piecewise objective functions, ii) the coupling among decision variables ($\alpha_{1},\alpha_{2}$)  in a complex non-linear way, iii) uncertain monotonicity and convexity due to the uncertain numerical values of parameters $Q_1,Q_2,c_1,c_2,Q_0,$ and $\epsilon_{0}$. We solve the complex Stage I-a through properly decomposing it into several subproblems, so that the objective functions can de divided into analyzable convex or monotonic pieces.

There are different market equilibria under different  conditions of sellers' marginal costs ($c_1,c_2$) and qualities ($Q_1,Q_2$). To capture these different cases,  we first introduce sellers' behavior patterns which heavily depend on sellers' marginal costs (Section \ref{behavior}). Then, we show how sellers' qualities affect the appearance of these behavior patterns (Section \ref{profile}).



\subsubsection{Behavior Pattern}
\label{behavior}
Behavior pattern specifies the two sellers' information disclosure decisions at the equilibrium  $(\alpha_1^*, \alpha_2^*)$.
\begin{lemma}
	\label{pattern}
	There exist four kinds of behavior patterns at the equilibrium:
	\begin{itemize}
		\item Pattern I: there is a unique equilibrium $(\alpha_1^*, \alpha_2^*)=(1,0)$.
		\item Pattern II:  there are two co-existing equilibria $(\alpha_1^*, \alpha_2^*)=(1,0)$ and $(\alpha_1^*, \alpha_2^*)=(0,1)$.
		\item Pattern III: there is a unique equilibrium  $(\alpha_1^*, \alpha_2^*)=(0,1)$.
		\item Pattern IV: there is a unique equilibrium  $(\alpha_1^*, \alpha_2^*)=(0,0)$. 
	\end{itemize}
\end{lemma}

Proof of Lemma \ref{pattern} is given in Appendix III. These behavior patterns largely depend on sellers' marginal costs ($c_1,c_2$). In Pattern I, seller 1  fully discloses information while seller 2 does not disclose information, which usually happens when seller 1 (the higher-quality seller) has a small marginal cost (as we shall see in Fig.~\ref{fig:t1t2} in Section \ref{profile}). In Pattern II,  one of the sellers disclosing information is an equilibrium, which happens when seller 1 has a moderate marginal cost and seller 2 has a small marginal cost. In Pattern III, seller 2  fully discloses information while seller 1 chooses no disclosure, which happens when seller 1 has a large marginal cost but seller 2's marginal cost is small. In Pattern IV, no seller discloses information, which usually happens when they both have high marginal costs.

Although different costs lead to different behavior patterns, some of the patterns do not exist under certain quality conditions ($Q_1,Q_2$). For example, as we shall see in Theorem \ref{type}, when both sellers have poor qualities, the equilibrium is $(\alpha_1^*, \alpha_2^*)=(0,0)$ (Pattern IV) regardless of the costs ($c_1,c_2$), and the other three patterns I, II, III are not achievable.
\subsubsection{Equilibrium Characterization}
\label{profile}
In the following, we  characterize the  equilibria by showing how the qualities affect the  behavior patterns.

\begin{figure*}
	\centering
	\subfigure[All-Achievable]{\label{fig:all}
		\begin{minipage}[t]{0.25\linewidth}
			\centering
			\includegraphics[width=1.5in]{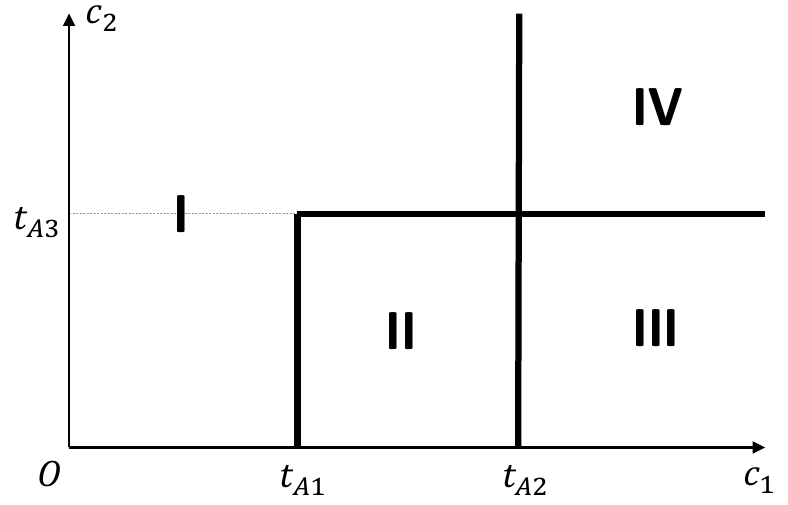}
		\end{minipage}%
	}%
	\subfigure[No-Mixed-Pattern]{\label{fig:no}
		\begin{minipage}[t]{0.25\linewidth}
			\centering
			\includegraphics[width=1.7in]{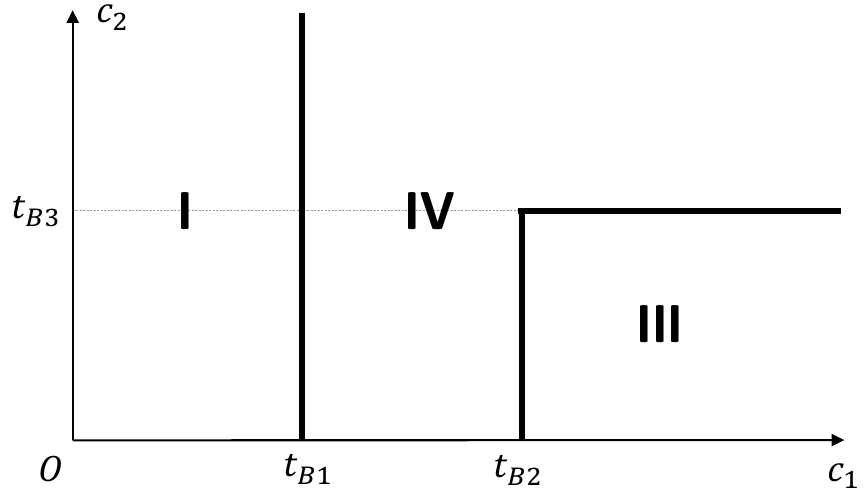}
		\end{minipage}%
	}%
	\subfigure[L-NeverDisclosure]{\label{fig:l}
		\begin{minipage}[t]{0.25\linewidth}
			\centering
			\includegraphics[width=1.5in]{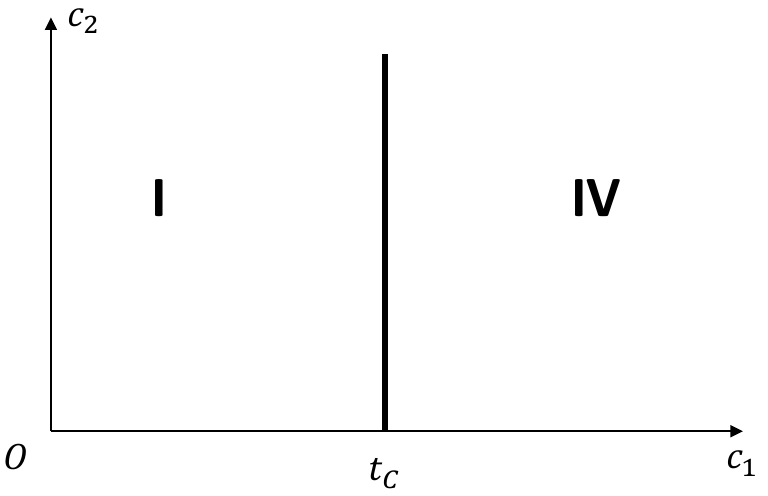}
		\end{minipage}%
	}%
	\subfigure[None-Disclosure]{\label{fig:none}
		\begin{minipage}[t]{0.25\linewidth}
			\centering
			\includegraphics[width=1.5in]{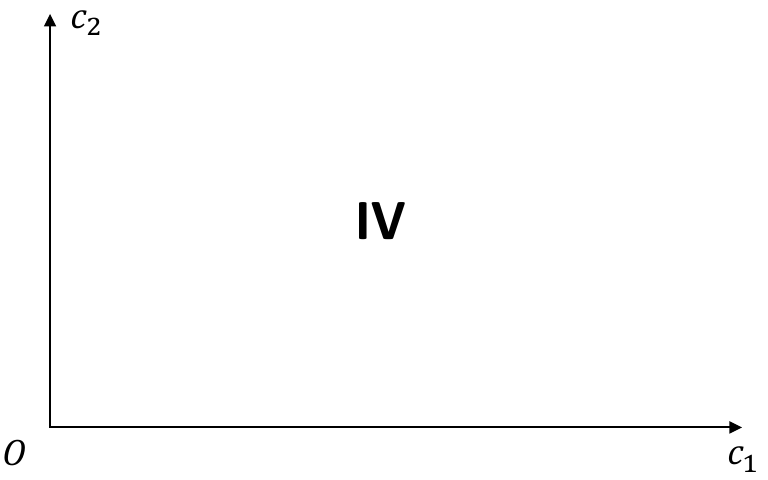}
		\end{minipage}%
	}%
	\vspace{-3mm}
	\caption{Equilibrium structures: equilibria under different costs ($c_1,c_2$).}
	\label{fig:t1t2}
	\vspace{-7mm}
\end{figure*}

\begin{figure}[tbp]
	\centering  	
	\includegraphics[width=0.75\linewidth]{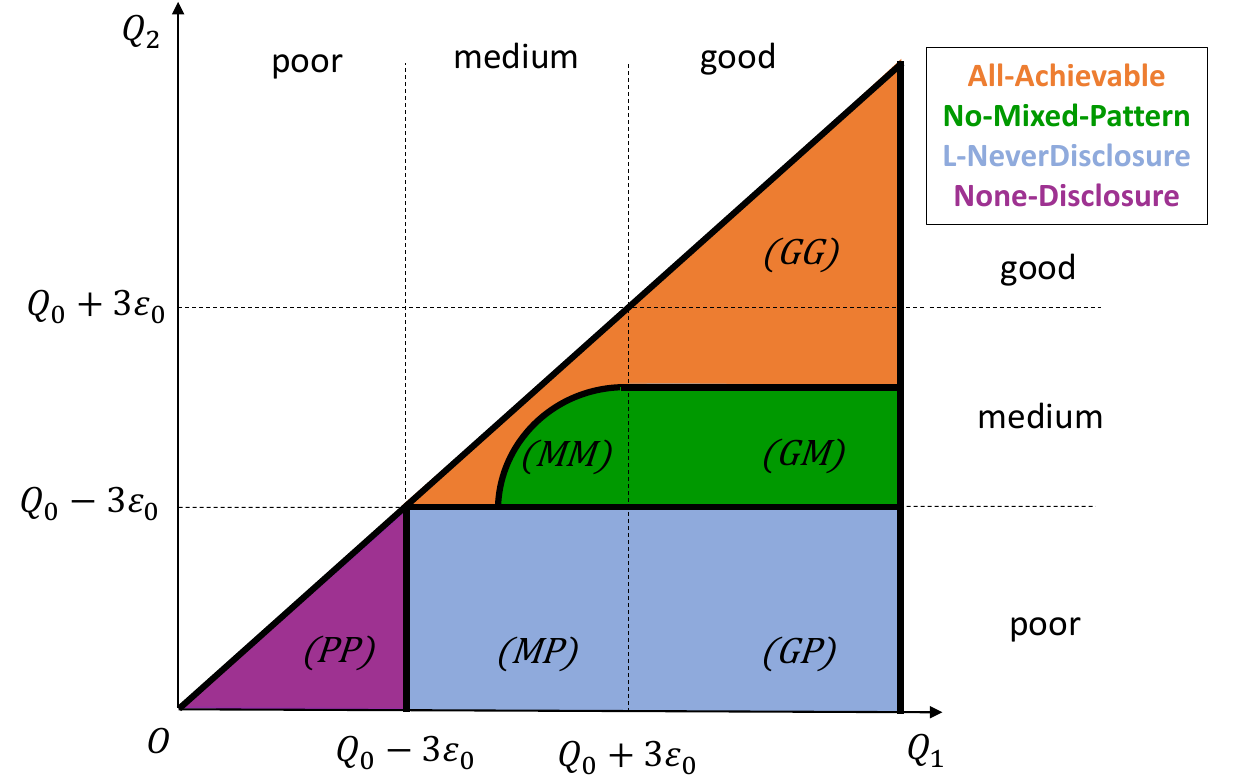}  
	\vspace{-3mm}
	\caption{Equilibrium structures under different qualities ($Q_1,Q_2$).} 
	\label{fig:q_6}  
	\vspace{-6mm} 
\end{figure}

\begin{theorem}
	\label{type}
	There exist four equilibrium structures, which are decided by sellers' qualities $(Q_1,Q_2)$:
	\begin{itemize}
		\item  ``All-Achievable": all patterns (I, II, III, IV) are achievable (Fig.~\ref{fig:t1t2}\subref{fig:all})\footnote{The notation $t_{A1}, t_{A2},t_{B1}, t_{B2}$, and $t_{C}$ in Fig.~\ref{fig:t1t2} are thresholds that distinguish different behavior patterns.			The specific values  vary with qualities and are summarized  in Appendix IV.}.  This happens when $(Q_1,Q_2)$ are in one of the following three cases:
		\begin{equation*}
		\begin{split}
		\text{1) } &Q_1\hspace{-0.5mm} \in\hspace{-0.5mm} \left[Q_0\hspace{-0.5mm}-\hspace{-0.5mm}3\epsilon_0, Q_0\hspace{-0.5mm}+\hspace{-0.5mm}(6\hspace{-0.5mm}-\hspace{-0.5mm}3\sqrt{5})\epsilon_0\right)\hspace{-0.5mm},Q_2\hspace{-0.5mm}\in \hspace{-0.5mm} \left[Q_0\hspace{-0.5mm}-\hspace{-0.5mm}3\epsilon_0,Q_1\right];\\
		\text{2) }& Q_1 \in \left[Q_0+(6-3\sqrt{5})\epsilon_0, Q_0+3\epsilon_0\right),\\
		&Q_2 \in  \left[Q_0-6\epsilon_0+\sqrt{54\epsilon_0^2-(Q_1-Q_0-6\epsilon_0)^2},Q_1\right];\\
		\text{3) }&Q_1\in \left[ Q_0+3\epsilon_0,+\infty\right), Q_2\in  \left[Q_0-6\epsilon_0+3\sqrt{5}\epsilon_0,Q_1\right].\\
		\end{split}
		\end{equation*}
		\item  ``No-Mixed-Pattern":  Patterns I, III and IV are achievable, but Pattern II (which contains two co-existing equilibria) is not achievable  (Fig.~\ref{fig:t1t2}\subref{fig:no}). This happens when  $(Q_1,Q_2)$ are in one of the following two cases:
		\begin{equation*}
		\begin{split}
		\text{1) }&Q_1\in  \left[Q_0+(6-3\sqrt{5})\epsilon_0,Q_0+3\epsilon_0\right),\\
		& Q_2\hspace{-0.5mm}\in\hspace{-0.5mm} \left[Q_0\hspace{-0.5mm}-\hspace{-0.5mm}3\epsilon_0,Q_0\hspace{-0.5mm}-\hspace{-0.5mm}6\epsilon_0\hspace{-0.5mm}+\hspace{-0.5mm} \sqrt{54\epsilon_0^2\hspace{-0.5mm}-\hspace{-0.5mm}(Q_1\hspace{-0.5mm}-\hspace{-0.5mm}Q_0\hspace{-0.5mm}-\hspace{-0.5mm}6\epsilon_0)^2}\right);\\
		\text{2) }& Q_1 \hspace{-0.5mm} \in \hspace{-0.5mm} \left[Q_0\hspace{-0.5mm}+\hspace{-0.5mm}3\epsilon_0,+\infty\right)\hspace{-0.5mm},\hspace{-0.5mm} Q_2 \hspace{-0.5mm} \in \hspace{-0.5mm} \left[Q_0\hspace{-0.5mm}-\hspace{-0.5mm}3\epsilon_0,\hspace{-0.5mm}Q_0\hspace{-0.5mm}-\hspace{-0.5mm}6\epsilon_0\hspace{-0.5mm}+\hspace{-0.5mm}3\sqrt{5}\epsilon_0\right)\hspace{-0.5mm}.\\
		\end{split}
		\end{equation*}
		\item   ``L-NeverDisclosure": only Patterns I and IV are achievable (Fig.~\ref{fig:t1t2}\subref{fig:l}). Seller 2 who has a lower quality never discloses information, and whether seller 1 discloses information depends on his cost. This happens when $0 \le Q_2<Q_0-3\epsilon_0<Q_1$.
		
		\item  ``None-Disclosure":   only Pattern IV is achievable  (Fig.~\ref{fig:t1t2}\subref{fig:none}). None of the sellers discloses information. This happens when $0 \le Q_2<Q_1<Q_0-3\epsilon_0$.
	\end{itemize}
	The relationship between qualities $(Q_1,Q_2)$ and the corresponding equilibrium structures is summarized  in Fig~\ref{fig:q_6}.
\end{theorem}

Proof of Theorem \ref{type} is given in Appendix III. For the convenience of presenting the insights, we define seller $i$'s commodity quality to be ``good ($G$)'' when $Q_i > Q_0 + 3\epsilon_0$, or ``medium ($M$)'' when $Q_0 - 3\epsilon_0<Q_i < Q_0 + 3\epsilon_0$, or ``poor ($P$)'' when $Q_i < Q_0 - 3\epsilon_0$. 
Thus, since $Q_1\ge Q_2$, there are  six possible quality profiles ($Q_1,Q_2$) in total: $GG$,  $GM$, $MM$, $GP$, $MP$, $PP$, which are also shown in Fig.~\ref{fig:q_6}.


First, as shown in Fig.~\ref{fig:q_6},  when qualities $(Q_1,Q_2)$ are given, the corresponding equilibrium structure belongs to one of the four cases. Second, as shown in Fig.~\ref{fig:t1t2}, given  the equilibrium structure, the equilibrium (i.e., behavior pattern) is uniquely determined by sellers costs $(c_1,c_2)$. Sellers tend to not disclose information when the disclosure  costs are unaffordable.   Third, Fig.~\ref{fig:t1t2} and Fig.~\ref{fig:q_6} show that  Pattern II only  happens when sellers' quality difference and costs difference are relatively small. 
Thus, sellers' strategies are more symmetric that either seller disclosing information is an equilibrium. 

We can further derive  the following insights about commodity qualities and sellers' disclosure strategies:
\begin{observation}
	\label{obs2}
	\begin{enumerate}
		\renewcommand{\labelenumi}{i)}
		\item A poor-quality seller will never disclose information,  even if the disclosure is costless.
		\renewcommand{\labelenumi}{ii)}
		\item If a  medium-quality seller discloses information at the equilibrium,  both sellers will have positive prices and positive expected profits.
		\renewcommand{\labelenumi}{iii)}
		\item If a good-quality seller  discloses information at the equilibrium,  the other seller will choose  a  zero price and obtain a zero profit.
	\end{enumerate}	
\end{observation}
We further elaborate the Observation \ref{obs2} in the following. (i) Such a non-disclosure situation exists in the profiles of $GP$, $MP$, and $PP$. This is surprising, as we would expect that a seller will disclose information when the privacy cost is zero.  The reason that this is not in general true is that when  the seller's quality is poor, disclosing information will decrease the buyer's payoff, which makes the seller less competitive. (ii) Since medium quality is close to the buyer's estimated quality,   both sellers have positive expected probabilities to win the buyer and  can expect positive profits. (iii) Since good quality is much higher than the buyer's estimated quality, if the seller who discloses information at equilibrium is of  good quality, he  will definitely win the competition and the other seller can only get a zero profit.

The results in this subsection show that information disclosure  decision significantly depends on  sellers' commodity qualities and privacy costs. Next, we will further  explore other influencing factors of information disclosure.

\vspace{-1mm}
\subsection{What Affects Information Disclosure?}
\vspace{-0.5mm}
\label{equilibrium}
According to  Sections \ref{customer}, \ref{price}, and  \ref{information}, we can summarize the equilibria of the two-stage game in Appendix V, including the sellers' information disclosure strategies in Stage I-a, sellers' pricing decisions in Stage I-b , and the buyer's decisions in Stage II. 
These results reflect the relationship between  price competition  and information disclosure (Section \ref{when}).  
In addition, the relationship between the buyer's estimation bias and  information disclosure is revealed in Section \ref{how}.
\subsubsection{Price Competition and Information Disclosure}
\label{when}
There are certain cases where  sellers do not disclose information, which reveal the relationship between  price competition  and information disclosure.
\begin{proposition}	
	Both sellers fully disclosing information is never  an equilibrium.
	\label{thm5}
\end{proposition}
It is natural to presume that  sellers with  good commodity qualities and low privacy costs will both choose to disclose information. However, Proposition \ref{thm5} shows that full disclosure by both sellers will never be an equilibrium. The best response for seller 2 should be no disclosure when seller 1 chooses to disclose information. This is because as  $Q_1\ge Q_2$, both sellers fully disclosing information will lead to severe market competition, under which seller 2 will choose a zero price and hence obtains a negative profit $-c_2$. As this is worse than no disclosure, hence we have the conclusion in Proposition \ref{thm5}.

Moreover, a seller does not disclose information when his cost is unaffordable or quality is poor. In this case the model degenerates to a pure price competition, as summarized in Proposition \ref{thm6}.  
\begin{proposition}
	When 
	none of the sellers choose to disclose information, the game degenerates into a simple pricing competition, in which  both sellers set the prices to zero and obtain a zero profit. 
	\label{thm6}
\end{proposition}

\vspace{-1mm}
\subsubsection{Estimation Bias and Information Disclosure}
\label{how}
To study the impact of the buyer's estimation bias on sellers' information disclosure, we further consider a benchmark case where the buyer has no estimation bias ($\epsilon_0=0$)  in Appendix VI. 
Proposition \ref{estimation} summarizes the comparison results:
\begin{proposition}
	\label{estimation}
	There are three key differences between the positive estimation bias ($\epsilon_0 > 0$) case and no estimation bias ($\epsilon_0 = 0$) case: 
	\begin{enumerate}
		\renewcommand{\labelenumi}{i)}
		\item The seller $i$ whose quality is below average ($Q_i<Q_0$) never discloses information when $\epsilon_0=0$; but he may do so when $\epsilon_0 > 0$.
		\renewcommand{\labelenumi}{ii)}
		\item The seller who does not disclose information earns a zero profit when $\epsilon_0=0$; but he may obtain  a positive profit when $\epsilon_0 > 0$.
		\renewcommand{\labelenumi}{iii)}
		\item It is never possible for both sellers to receive positive expected profits when $\epsilon_0=0$. However, it is possible when $\epsilon_0 > 0$.
	\end{enumerate}
\end{proposition}
Proposition \ref{estimation} shows that the buyer's estimation bias   mitigates  market competition and encourages  information disclosure. 
Specifically, (i)  the estimation bias motivates the seller whose commodity quality is below average to disclose information;  (ii) the estimation bias enables sellers     who do not disclose information to achieve positive expected profits; (iii)  the estimation bias makes it possible for both sellers  to receive positive expected  profits. Therefore, when there is an increasing heterogeneity of buyers in the system (and thus diverse estimation biases), e.g., when the system expands buyer's market, there will be more disclosure and less competition.

As a quick summary, we have shown that  full disclosure  or non-disclosure by both sellers will both cause intense  price competition, but the existence of estimation biases can mitigate this effect. Next, we  study whether above results still hold when the system consists of multiple users, and analyze how sellers' capacities  affect market competition and  information disclosure.

\vspace{-2mm}
\section{General Setting: $N$ Sellers and $K$ Buyers}
\vspace{-1mm}
\label{multiple}
In this section, we consider the more general case of a set $\mathcal{N}=\{1,2, ..., N\}$ sellers and $\mathcal{K}=\{1, 2, ..., K\}$ buyers.  
To reveal the  impact of sellers' capacities on  information disclosure, we further analyze the equilibria  where sellers have unlimited or limited capacities,  respectively.

We still adopt the two-stage game model. Sellers make their information disclosure decisions    and prices in Stage I. In Stage II, buyers come to the system one by one, and each buyer purchases  one  commodity from a seller with a spare capacity.

We make several assumptions. First,    different buyers have different estimation biases, which are i.i.d., i.e., following the same distribution $\epsilon_k \sim U(-\epsilon_0,\epsilon_0),\forall k \in \mathcal{K}$.  Second, 
since solving the non-convex optimization problem in the basic setting is already challenging, it's quite difficult to perform pairwise comparison for $N$ sellers and then  derive all sellers' pricing strategies as well as  information disclosure levels. We can observe that  although sellers' strategy space of information disclosure level is $[0,1]$ in the basic setting, the result at the equilibrium turns out to be either 1 or 0 (Lemma \ref{pattern}).  Thus, to simplify the analysis, we assume that sellers' information disclosure strategy space is $\{0,1\}$ in this section. 

In this case, the $N$ Sellers' Information Disclosure Game and the corresponding equilibrium are\footnote{As we mainly focus on  studying sellers' information disclosure, the analysis of buyers' optimal strategies and sellers' optimal  prices in this section are given in Appendices VII-X.}:
\begin{defn}[Stage I-a: $N$ Sellers' Information Disclosure Game]\quad \\
	\vspace{-5mm}
	\begin{itemize}
		\item Players:  sellers in $\mathcal{N}$.
		\item Strategy space: each seller $i \in \mathcal{N}$ chooses his information disclosure level $\alpha_i \in  \{0,1\}$.
		\item Payoff function: seller $i$  maximizes his expected profit $E[W_i(\alpha_i;\boldsymbol{\alpha_{-i}})]$, where $\boldsymbol{\alpha_{-i}} \triangleq (\alpha_k,\forall k \in \mathcal{N},k\neq i)$.
	\end{itemize}
\end{defn}

\begin{defn}
	The equilibrium of the $N$ Sellers' Information Disclosure Game is a profile $\boldsymbol{\alpha^*}=(\alpha_1^*,...,\alpha_N^*) \subseteq \{0,1\}^N$ such that for each seller $i\in \mathcal{N}$,
	\begin{equation*}
	E[W_i(\alpha_i^*,\boldsymbol{\alpha_{-i}^*})] \ge E[W_i(\alpha_i,\boldsymbol{\alpha_{-i}^*})], \forall \alpha_i \in \{0,1\}.
	\end{equation*}
\end{defn}

In the following, we will consider  three possible scenarios  of sellers' capacities (i.e., the number of buyers that sellers can serve).
\begin{enumerate}
	\renewcommand{\labelenumi}{i)}
	\item Each seller has an \emph{unlimited capacity}. This is  a reasonable approximation of the practical case where each seller  has an enough capacity   to serve all the coming buyers. For example, some sellers in an accommodation system own many properties or chain hotels.
	\renewcommand{\labelenumi}{ii)}
	\item Each seller has a \emph{single capacity (one buyer per seller)}. This is motivated by the practical case where a worker in labor sharing systems can only serve one buyer at a time.
	\renewcommand{\labelenumi}{iii)}
	\item Each seller has a \emph{limited capacity (multiple buyers per seller)}. This corresponds to the case where a seller can sell his leftover network resource to multiple buyers at the same time. The first and second scenarios can be regarded as  special cases of this more general scenario. 
\end{enumerate}

\vspace{-4mm}
\subsection{Unlimited Capacity}
\vspace{-1mm}
\label{U}
In this subsection, we analyze sellers' information disclosure strategies under the scenario where each seller has an unlimited capacity. The analysis also serves as the benchmark for the limited capacity scenarios.



We denote by $\Phi_{U}^e$  the set of sellers who disclose information simultaneously  at the  equilibrium $e_{U}$, where $e_{U}$ can be any possible equilibrium in the unlimited capacity scenario.  Denote by $|\Phi_{U}^e|$  the number  of sellers who disclose information at the equilibrium $e_{U}$. Given the commodity qualities and privacy costs of sellers, there can be multiple co-existing equilibria. Theorem \ref{lll} shows that there are only two types of equilibria: 
\begin{theorem}
	\label{lll}
	In the unlimited capacity scenario, the equilibrium always exists but may not be unique. There exist two kinds of equilibria:
	\begin{enumerate}
		\renewcommand{\labelenumi}{i)}
		\item One-seller disclosure (i.e., $|\Phi_{U}^e|=1$) is the equilibrium  if and only if   at least one seller $i\in \mathcal{N}$ has  $(Q_i,c_i)$ in one of the two cases in \eqref{important}. 
		\vspace{-1mm}
		\begin{equation}
		\label{important}
		\begin{split}
		&\text{1) }Q_i    \in   \left(Q_0+ \epsilon_0,  Q_0+ 3\epsilon_0\right),c_i \in  \left[0,\frac{(Q_i-Q_0+\epsilon_0)^2}{8\epsilon_0}\right];\\
		&\text{2) }Q_i \in \left[  Q_0+ 3\epsilon_0,\infty\right),c_i \in \left[0,Q_i-Q_0-\epsilon_0\right].
		\end{split}
		\end{equation}
		
		The only  seller $i$ who discloses information at the equilibrium (i.e., $\alpha_i^*=1,\boldsymbol{\alpha_{-i}^*}=\boldsymbol{0}$) has  $(Q_i,c_i)$ satisfying   \eqref{important} and \eqref{10},
		\vspace{-1mm}
		\begin{equation}
		\label{10}
		Q_i \ge Q_j-c_j,\forall j \in \mathcal{N},j\neq i.
		\end{equation}
		The equilibrium is not unique when more than one seller has  $(Q_i,c_i)$ satisfying \eqref{important} and  \eqref{10}. Any such seller disclosing information while other sellers not is an equilibrium.
		\renewcommand{\labelenumi}{ii)}
		\item No disclosure is the equilibrium (i.e., $|\Phi_{U}^e|=0$) if and only if 
		$(Q_i,c_i)$  does not satisfy \eqref{important}, for each seller $ i \in \mathcal{N}$.
	\end{enumerate}	
\end{theorem}
Proof of Theorem \ref{lll} is given in Appendix VIII.  Theorem \ref{lll}  shows that at most one seller discloses information at an equilibrium in the unlimited capacity scenario (i.e., $|\Phi_{U}^e|\le 1$), which is a bit counter-intuitive. If  more than one seller discloses information, all buyers will choose the seller who gives the highest payoff since the seller has an unlimited capacity. Thus, those unselected sellers who have  disclosed information will obtain negative profits.
It is impossible for two sellers who disclose information  at the equilibrium to provide the same highest payoff for buyers, even if they have the same  quality. A slight price reduction of one of the two sellers will attract all buyers, so they will keep decreasing their prices in the competition until one seller's profit is  no larger than no disclosure.  
Hence, multiple sellers disclosing information  cannot be an equilibrium in the unlimited capacity scenario. This is consistent with the basic setting (Proposition \ref{thm5}).

\begin{figure}[tbp]
	\centering  	
	\includegraphics[width=0.4\linewidth]{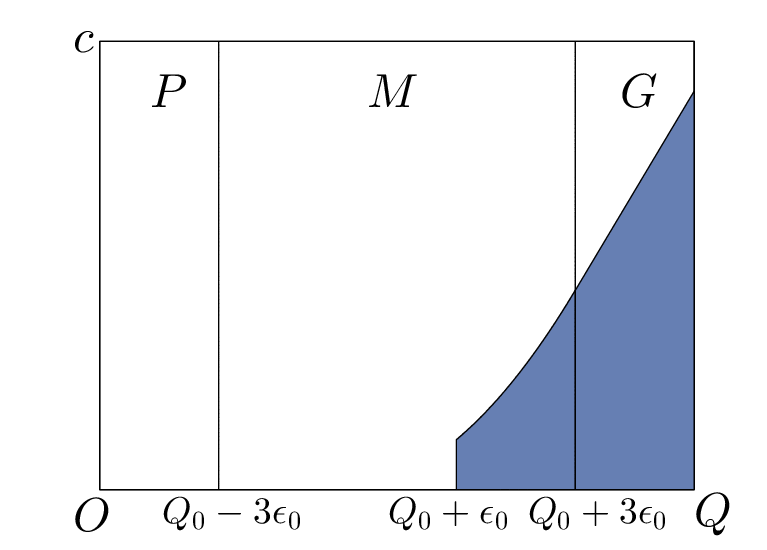}  
	\vspace{-3mm}
	\caption{Illustration of \eqref{important}:   each seller with ($Q,c$) in the blue region is willing to disclose information when others do not.} 
	\label{dgd}  
	\vspace{-5mm} 
\end{figure}

More importantly, Theorem \ref{lll}  specifies which seller  discloses information at an equilibrium   in the unlimited capacity scenario. The condition   \eqref{important} means that seller $i$ is willing to disclose information when others do not. From the illustration of \eqref{important} in Fig.~\ref{dgd}, we can observe that the sellers who are willing to disclose information have high qualities and low costs. The condition \eqref{10} ensures that seller $j$ will not disclose information when seller $i$ does, which requires the seller $i$ who fully discloses his information  at an equilibrium  to have a  quality  substantially better than all other sellers.


Next, we study the impact of capacities on sellers' information disclosure  by considering the limited capacity scenario.

\vspace{-2mm}
\subsection{Single Capacity (One Buyer Per Seller)}
\vspace{-1mm}
\label{LO}
In this subsection, we focus on the sellers' information disclosure strategies when each seller has a limited capacity and can only serve one buyer. 

Since  we are studying sellers' strategies under competition, we assume that  supply is larger than demand\footnote{The analysis of the scenario where supply is no larger than demand (i.e., no game among sellers) is given in Appendix XII.}.   
We denote by $\Phi_{SO}^e$  the set of sellers who disclose information simultaneously  at the equilibrium $e_{SO}$, where $e_{SO}$ can be any possible equilibrium in the single capacity (one buyer per seller) scenario\footnote{Subscript $SO$ represents single capacity (one buyer per seller), similarly in the following, subscript $U$ represents unlimited capacity and subscript $LM$ represents limited capacity (multiple buyer per seller).}. Denote by $|\Phi_{SO}^e|$  the number  of sellers who disclose information at the equilibrium $e_{SO}$. Given the  qualities and costs of sellers, there can be multiple co-existing equilibria. Theorem \ref{sfd} shows that there are only two types of equilibria:
\begin{theorem}
	\label{sfd}
	In the single capacity (one buyer per seller) scenario,  the equilibrium always exists but may not be unique. There exist two kinds of equilibria:
	\begin{enumerate}
		\renewcommand{\labelenumi}{i)}
		\item At the equilibrium, $1\le  |\Phi_{SO}^e|\le K$  if and only if at least one seller $i\in \mathcal{N}$ has 		$(Q_i,c_i)$  satisfying \eqref{important}.
		
		Each seller $i \in   \Phi_{SO}^e$ who discloses information at the equilibrium has  $(Q_i,c_i)$ satisfying both \eqref{important} and \eqref{11},
		\begin{equation}
		\label{11}
		Q_i\ge  Q_j-c_j, \forall i\in \Phi_{SO}^e,\forall j\in \mathcal{N}\backslash \Phi_{SO}^e.
		\end{equation}
		The equilibrium is not unique when more than $K$ sellers have  $(Q_i,c_i)$ satisfying \eqref{important} and  \eqref{11}. Any $K$ of them disclosing information while other sellers not is an equilibrium.
		\renewcommand{\labelenumi}{ii)}
		\item At the equilibrium,  $ |\Phi_{SO}^e|=0$  if and only if 
		$(Q_i,c_i)$  does not satisfy \eqref{important}, for each seller  $i \in \mathcal{N}$.
	\end{enumerate}
	%
	%
\end{theorem}
Proof of Theorem \ref{sfd} is given in Appendix IX. Theorem \ref{sfd} shows that at most $K$ sellers fully disclose information at an equilibrium in this scenario, where $K$ is the number of buyers. Each seller can  serve one of the $K$ buyers. If there are more than $K$ sellers with full disclosure, the unselected sellers  will get  negative profits, which is worse than no disclosure. Hence, this will not happen at the equilibrium.  

Moreover, Theorem \ref{sfd}  identifies which sellers  disclose information at an equilibrium in the single capacity (one buyer per seller) scenario. The results show that, in case (i), there is  competition among the sellers,  and those who eventually disclose information at an equilibrium need to have  qualities significantly better than those sellers who choose not to disclose information  (i.e., \eqref{11}). In case (ii),  for any seller, disclosure leads to  a lower profit than non-disclosure, so no seller will disclose information. 

%


Finally, we will consider the most general case where   each seller can serve multiple buyers.

\vspace{-2mm}
\subsection{Limited Capacity (Multiple Buyers Per Seller)}
\vspace{-1mm}
\label{HL}
In this subsection, we analyze the sellers' information disclosure strategies under the scenario where  sellers  have limited capacities and each seller can serve multiple buyers. 

Let $\omega_i \in (1,+\infty)$ denote the capacity of seller $i$. If $ \omega_i = \infty, \forall i \in \mathcal{N}$, it corresponds  to the unlimited capacity scenario; If $ \omega_i = 1, \forall i \in \mathcal{N}$, it corresponds  to the single capacity (one buyer per seller) scenario. Since we are studying sellers' behaviors in competition, we still assume that  supply is larger than demand in this subsection. 
We denote by $\Phi_{LM}^e$  the set of sellers who disclose information at the  equilibrium $e_{LM}$, where $e_{LM}$ can be any possible equilibrium in the limited capacity (multiple buyers per seller) scenario.  Denote by $|\Phi_{LM}^e|$  the number  of sellers who disclose information at the equilibrium $e_{LM}$. Given the  qualities and costs of sellers, there can be multiple co-existing equilibria. Theorem \ref{jfk} shows that there are only two types of equilibria:
\begin{theorem}
	\label{jfk}
	In the limited capacity (multiple buyers per seller) scenario,  the equilibrium always exists but may not be unique. There exist two kinds of equilibria:
	\begin{enumerate}
		\renewcommand{\labelenumi}{i)}
		\item 
		At the equilibrium,  $1\le |\Phi_{LM}^e|<K$ 
		if and only if at least one seller $i\in \mathcal{N}$ has $(Q_i,c_i)$  satisfying \eqref{important}.
		
		Each seller $i \in   \Phi_{LM}^e$ who discloses information at equilibrium has  $(Q_i,c_i)$ satisfying both \eqref{important} and \eqref{12}.
		\begin{equation}
		\label{12}
		Q_i\ge  Q_j-c_j,\forall i\in \Phi_{LM}^e,\forall j\in \mathcal{N}\backslash\Phi_{LM}^e
		\end{equation}
		The equilibrium is not unique when the total capacity of a proper subset of sellers with  $(Q_i,c_i)$ satisfying \eqref{important} and  \eqref{12} is larger than or equal to $K$. Any subset of these sellers with  capacity  satisfying \eqref{a} disclosing information while other sellers not is an equilibrium.
		\vspace{-1mm}
		\begin{subequations}
			\label{a}
			\begin{align}
			\sum_{i \in \Phi_{LM}^e}\omega_i &\ge K,\label{11a}\\
			\sum_{i \in \Phi_{LM}^e}\omega_i -\omega_j&<K,\; j=\arg\min_{i\in \Phi_{LM}^e}\{Q_i-p_i\}.\label{11b}
			\end{align}
		\end{subequations}
		\renewcommand{\labelenumi}{ii)}
		\item At the equilibrium, $|\Phi_{LM}^e|=0$  if and only if 
		$(Q_i,c_i)$  does not satisfy \eqref{important}, for each seller $ i \in \mathcal{N}$.
	\end{enumerate}	
\end{theorem}

\begin{figure*}
	\centering
	\subfigure[Single Capacity Scenario ($\omega=1$)]{ \label{fig:c}
		\begin{minipage}[t]{0.311\linewidth}
			\centering
			\includegraphics[width=1.4in]{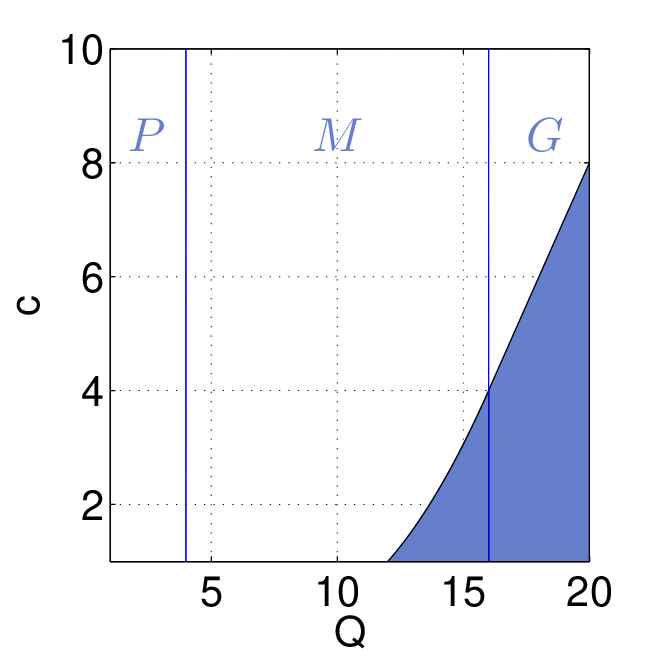}
		\end{minipage}%
	}%
	\;
	\subfigure[Limited Capacity Scenario ($\omega=8$)]{ \label{fig:b}
		\begin{minipage}[t]{0.311\linewidth}
			\centering
			\includegraphics[width=1.4in]{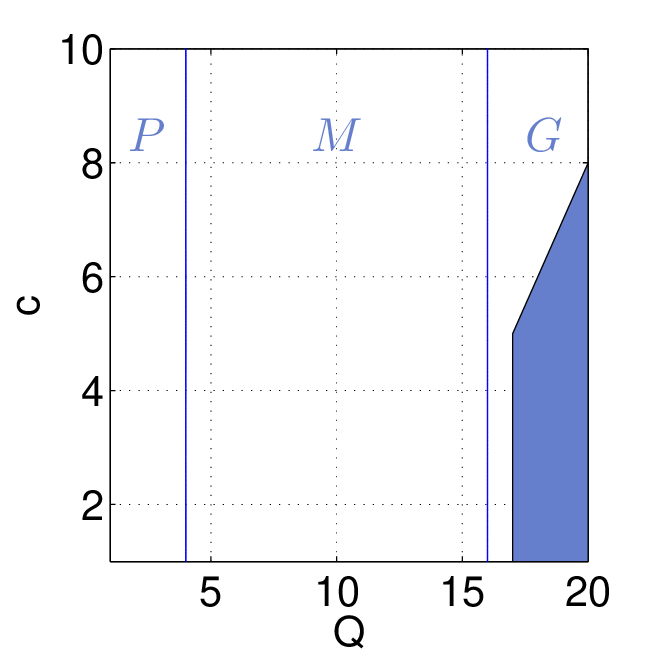}
		\end{minipage}%
	}%
	\;
	\subfigure[Unlimited Capacity Scenario ($\omega=\infty$)]{ \label{fig:a}
		\begin{minipage}[t]{0.311\linewidth}
			\centering
			\includegraphics[width=1.4in]{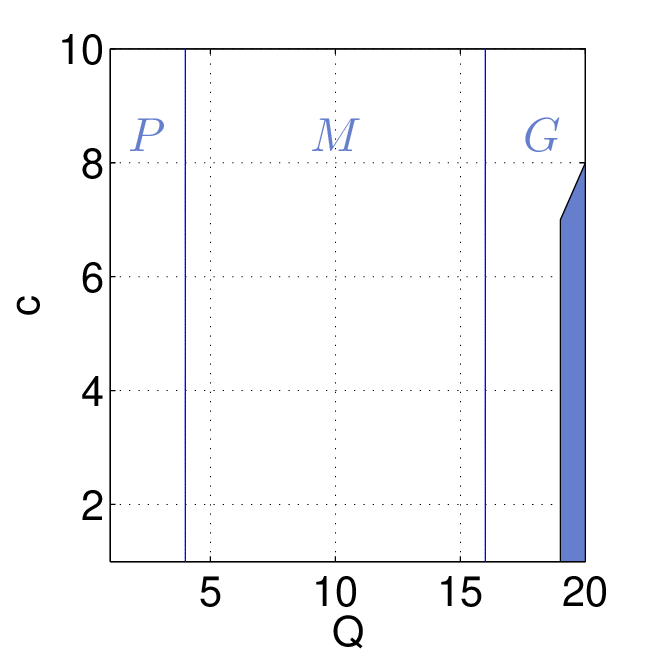}
		\end{minipage}%
	}%
	\vspace{-2mm}
	\caption{Eligible region: sellers with ($Q,c$) in the blue region can disclose information at the  equilibria.}
	\label{normal}
	\vspace{-6mm}
\end{figure*}

Proof of Theorem \ref{jfk} is given in Appendix X. Theorem \ref{jfk} characterizes the equilibria in the limited capacity (multiple buyers per seller) scenario and  identifies which sellers disclose information at an equilibrium. 
In case (i),   if the total capacity of sellers who are eligible to disclose information (i.e., satisfying \eqref{important} and  \eqref{12}) is larger than the number of buyers $K$,   the sellers who eventually disclose information at an equilibrium (i.e., sellers in $\Phi_{LM}^e$) will have exactly enough capacity for buyers. To be more specific, equation \eqref{11a} means that  total capacity of  the  sellers in $\Phi_{LM}^e$ is no smaller than $K$. Equation \eqref{11b} means that  at the equilibrium,  without the seller who offers buyers the smallest payoff   among the   sellers in $\Phi_{LM}^e$,  the  remaining sellers in $\Phi_{LM}^e$ cannot  serve all $K$ buyers. 
One may presume that some sellers in $\Phi_{LM}^e$ may have some unsold capacity such that \eqref{11b}  may not hold. However,  it is impossible that there exist two sellers who disclose information and both have idle capacities, even if they have the same  quality. Buyers  choose sellers who offer higher payoffs to them, so two sellers having idle capacities means that they provide the same lowest payoff among disclosure sellers (i.e., $\min_{i\in \Phi^e_{LM}}\{Q_i-p_i\}$) to buyers. A slight price reduction of one of the two sellers will attract the buyers who have chosen the other seller, so they will keep decreasing their prices in the competition until a seller's profit is  no larger than no disclosure. 
In case (ii), no seller is willing to disclose information as  all sellers have  low qualities or high costs (i.e., not satisfy \eqref{important}). 

Next, we will compare the number of sellers who disclose information at equilibria in three capacity scenarios.

\vspace{-3mm}
\subsection{Comparison of Information Disclosure in Three Scenarios}
\vspace{-1mm}
In this subsection, we compare the results derived in the three capacity scenarios to explore the impact of capacity limitation on information disclosure. 

Recall that $e_U$, $e_{SO}$ and $e_{LM}$ denote the equilibria under the three  capacity scenarios, respectively. Moreover,    $|\Phi_U^e|$, $|\Phi_{SO}^e|$, and $|\Phi_{LM}^e|$ are  the corresponding numbers of sellers who disclose information at the equilibria. 
Based on  Theorem \ref{lll}, Propositions \ref{sfd} and \ref{jfk}, we have the following relationship between capacity limitation and information disclosure:
\begin{corollary}
	\label{obbb}
	For the same set of sellers and buyers, given any  $e_U$, $e_{SO}$ and $e_{LM}$, the number of sellers who disclose information at the equilibrium satisfies 	$|\Phi_U^e| \le |\Phi_{LM}^e| \le |\Phi_{SO}^e|$.
\end{corollary}
Proof of Corollary \ref{obbb} is given in Appendix XI. Corollary \ref{obbb} shows that more limited capacity   allows more sellers to disclose information, encouraging the information disclosure in the system. Moreover,  non-disclosure sellers will obtain a zero profit due to the fierce competition, so capacity limitation also enables more sellers to get positive profits, mitigating the market competition.

\vspace{-2mm}
\section{Numerical Analysis}
\vspace{-1mm}
In this section, we conduct numerical experiments  to validate our analytical results and insights.

So far, we have derived the closed-form solutions to the equilibria under three capacity scenarios (Theorems \ref{lll}, \ref{sfd}, and \ref{jfk}), and we  have rigorously proved the relationship between sellers' capacities and information disclosure (Corollary \ref{obbb}).  

Next, we will give a concrete illustration through simulation.
Since we study the competition among sellers, we set the number of sellers $N=200$, which  is larger than the number of buyers $K=32$. As we assume that   sellers' commodity qualities $Q$ takes the values in $[1,20]$,  we set buyers' average prior believe $Q_0$ to be $10$, which is  the average commodity quality in the market. The estimation bias is $\epsilon_0=2$. The marginal privacy cost $c$  in $[1,10]$ can   capture different cases where sellers with high, medium, or low privacy concerns. 

As shown in Fig.~\ref{normal}, we simulate  three capacity scenarios, respectively. More specifically,   $\omega=1$ is the single capacity (one buyer per seller) scenario,   $\omega=8$ is   the limited capacity (multiple buyers per seller) scenario, and    $\omega=\infty$ is the unlimited capacity scenario.  
The x-axis is the commodity quality and y-axis is the marginal privacy cost. 
The notations of ``$P$''(poor),``$M$''(medium) and ``$G$''(good)  refer to the partition of sellers' qualities, which have been defined in Section \ref{profile}.  
In each of the subfigures, the blue shaded area corresponds to the eligible region, within which each seller can be the seller who discloses information at equilibria. For example,  any seller  in the eligible region of Fig.~\ref{normal}\subref{fig:a} disclosing information and other sellers not is an equilibrium  in the unlimited capacity scenario. 

Fig.~\ref{normal}  shows that the sellers who can disclose information (i.e., in the eligible region) have high qualities and low marginal costs (as proved in  Theorems \ref{lll}, \ref{sfd}, and \ref{jfk}).   
Furthermore,  larger capacities (i.e., larger $\omega$) require the sellers with full disclosure at an equilibrium to have higher (more competitive) qualities, corresponding to $Q\ge 12$ in single capacity scenario, $Q\ge 17$ in limited capacity scenario, and $Q\ge 19$ in unlimited capacity scenario in Fig.~\ref{normal}. Therefore, as the sellers' capacities increase,  a smaller number of sellers  can disclose information at an equilibrium. This is consistent with our finding that larger capacity strengthens the competition among sellers and thus discourages the information disclosure.

Moreover, this paper provides a clear understanding of the observations   in previous numerical studies in  literature and some system regulations in practice.  For example, Xu \emph{et al.} \cite{xu2021impact} numerically showed that extra information may exhibit a diminishing effect on buyer purchase behavior. This has been validated and explained by our paper that disclosing information is worse than no disclosure in some cases (Theorem \ref{type}).  Airbnb has launched the Airbnb Select program to improve property qualities  and has restricted hosts from renting out multiple properties in some areas \cite{tim,qtimmurphy.org}. Both regulations  eventually encourage sellers' information disclosure according to our analysis (Theorem \ref{type} and Corollary \ref{obbb}).  

Overall, the results in this paper  provide guidelines for sellers to optimize information disclosure behaviors in sharing systems. The results can also facilitate sharing systems' regulation decisions and  incentive mechanism design.

\section{Conclusion}
\label{conclusion}
This paper  studied  sellers' information disclosure strategies in competition in sharing systems. We   captured the trade-off between sellers' commodity display effects and privacy costs, and   proposed a  two-stage game to characterize the interactions between sellers and buyers. We   completely presented  the complex structure of equilibria under different conditions despite the non-convexity of the optimization problem. The results showed  that full disclosure by all sellers will never be an equilibrium even if they all have good qualities and low costs; none of the sellers disclosing information will make all sellers get zero profit. 
In the future, we will analyze the general case where sellers can choose to disclose information in a much finer granularity (even in the continuous fashion).  

\vspace{-1mm}
\bibliographystyle{IEEEtran}
\bibliography{ref}

\begin{thebibliography}{10}
\providecommand{\url}[1]{#1}
\csname url@samestyle\endcsname
\providecommand{\newblock}{\relax}
\providecommand{\bibinfo}[2]{#2}
\providecommand{\BIBentrySTDinterwordspacing}{\spaceskip=0pt\relax}
\providecommand{\BIBentryALTinterwordstretchfactor}{4}
\providecommand{\BIBentryALTinterwordspacing}{\spaceskip=\fontdimen2\font plus
\BIBentryALTinterwordstretchfactor\fontdimen3\font minus
  \fontdimen4\font\relax}
\providecommand{\BIBforeignlanguage}[2]{{%
\expandafter\ifx\csname l@#1\endcsname\relax
\typeout{** WARNING: IEEEtran.bst: No hyphenation pattern has been}%
\typeout{** loaded for the language `#1'. Using the pattern for}%
\typeout{** the default language instead.}%
\else
\language=\csname l@#1\endcsname
\fi
#2}}
\providecommand{\BIBdecl}{\relax}
\BIBdecl

\bibitem{ningglobe}
N.~Ding, Z.~Fang, and J.~Huang, ``Information disclosure game on sharing
  platforms,'' in \emph{IEEE Global Communications Conference (GLOBECOM)},
  2020.

\bibitem{Yu2020}
J.~Yu, M.~H. Cheung, and J.~Huang, ``Economics of mobile data trading market,''
  \emph{{IEEE} Transactions on Mobile Computing}, 2020.

\bibitem{2cm}
``{2CM},'' \url{https://www.hk.chinamobile.com/tc/pop_ups/2cm_tnc.html}.

\bibitem{Khalili2015}
M.~M. Khalili, L.~Gao, J.~Huang, and B.~H. Khalaj, ``Incentive design and
  market evolution of mobile user-provided networks,'' in \emph{{IEEE}
  Conference on Computer Communications Workshops}, 2015.

\bibitem{bewifi}
``{BeWiFi},'' \url{https://bewifi.co/}.

\bibitem{airbnb}
``Airbnb,'' \url{https://www.airbnb.com/}.

\bibitem{house}
``House{T}rip,'' \url{https://www.housetrip.com/}.

\bibitem{task}
``Task{R}abbit,'' \url{https://www.taskrabbit.com/}.

\bibitem{Hamari2015}
J.~Hamari, M.~Sjöklint, and A.~Ukkonen, ``The sharing economy: Why people
  participate in collaborative consumption,'' \emph{Journal of the Association
  for Information Science and Technology}, vol.~67, no.~9, pp. 2047--2059,
  2015.

\bibitem{Singh2018}
V.~Singh, A.~Bhattacherjee, J.~Kalapatapu, U.~Srivastava, and S.~Fu, ``The
  effect of image choice on airbnb reservations: A combination of deep learning
  and econometric analysis,'' \emph{Americas conference on information
  systems}, 2018.

\bibitem{Balau2016}
N.~B{\u{a}}l{\u{a}}u and S.~Utz, ``Information sharing as strategic behaviour:
  the role of information display, social motivation and time pressure,''
  \emph{Behaviour {\&} Information Technology}, vol.~36, no.~6, pp. 589--605,
  2016.

\bibitem{Liang2019}
S.~Liang, H.~Li, X.~Liu, and M.~Schuckert, ``Motivators behind information
  disclosure: Evidence from airbnb hosts,'' \emph{Annals of Tourism Research},
  vol.~76, pp. 305--319, 2019.

\bibitem{Lutz2017}
C.~Lutz, C.~P. Hoffmann, E.~Bucher, and C.~Fieseler, ``The role of privacy
  concerns in the sharing economy,'' \emph{Information, Communication {\&}
  Society}, vol.~21, no.~10, pp. 1472--1492, 2017.

\bibitem{Ma2017}
Q.~Ma, J.~Huang, T.~Ba{\c{s}}ar, J.~Liu, and X.~Chen, ``Pricing for sharing
  economy with reputation,'' \emph{{ACM} {SIGMETRICS} Performance Evaluation
  Review}, vol.~44, no.~3, pp. 32--32, 2017.

\bibitem{Zervas2015}
G.~Zervas, D.~Proserpio, and J.~Byers, ``A first look at online reputation on
  airbnb, where every stay is above average,'' \emph{{SSRN} Electronic
  Journal}, 2015.

\bibitem{Fradkin2018}
A.~Fradkin, E.~Grewal, and D.~Holtz, ``The determinants of online review
  informativeness: Evidence from field experiments on airbnb,'' \emph{{SSRN}
  Electronic Journal}, 2018.

\bibitem{Kim2015WhyPP}
J.~Kim, Y.~Yoon, and H.~Zo, ``Why people participate in the sharing economy: A
  social exchange perspective,'' in \emph{PACIS}, 2015.

\bibitem{xie2017impacts}
K.~Xie and Z.~Mao, ``The impacts of quality and quantity attributes of airbnb
  hosts on listing performance,'' \emph{International Journal of Contemporary
  Hospitality Management}, vol.~29, no.~9, pp. 2240--2260, 2017.

\bibitem{tim}
D.~Ting and Skift, ``Airbnb is set to launch a new tier of select properties,''
  \url{https://skift.com/2018/02/13/airbnb-is-set-to-launch-a-new-tier-of-select-properties/}.

\bibitem{qtimmurphy.org}
M.~Nickelsburg, ``Seattle approves new airbnb regulations to limit short-term
  rentals to 2 units per host,''
  \url{https://www.geekwire.com/2017/seattle-approves-new-airbnb-regulations-limit-short-term-rentals}.

\bibitem{nie2020multi}
J.~Nie, J.~Luo, Z.~Xiong, D.~Niyato, P.~Wang, and H.~V. Poor, ``A multi-leader
  multi-follower game-based analysis for incentive mechanisms in socially-aware
  mobile crowdsensing,'' \emph{IEEE Transactions on Wireless Communications},
  vol.~20, no.~3, pp. 1457--1471, 2020.

\bibitem{sedghani2021incentive}
H.~Sedghani, D.~Ardagna, M.~Passacantando, M.~Z. Lighvan, and H.~S. Aghdasi,
  ``An incentive mechanism based on a stackelberg game for mobile crowdsensing
  systems with budget constraint,'' \emph{Ad Hoc Networks}, vol. 123, p.
  102626, 2021.

\bibitem{Zheng2020}
Z.~Zheng, Y.~Peng, F.~Wu, S.~Tang, and G.~Chen, ``{ARETE}: On designing joint
  online pricing and reward sharing mechanisms for mobile data markets,''
  \emph{{IEEE} Transactions on Mobile Computing}, vol.~19, no.~4, pp. 769--787,
  2020.

\bibitem{Lu2020}
Y.~Lu, Y.~Qi, S.~Qi, Y.~Li, H.~Song, and Y.~Liu, ``Say no to price
  discrimination: Decentralized and automated incentives for price auditing in
  ride-hailing services,'' \emph{{IEEE} Transactions on Mobile Computing},
  2020.

\bibitem{Courcoubetis2019}
C.~Courcoubetis and A.~Dimakis, ``Throughput and pricing of ridesharing
  systems,'' in \emph{{IEEE} Conference on Computer Communications}, 2019.

\bibitem{Lin2019}
Q.~Lin, W.~Xu, M.~Chen, and X.~Lin, ``A probabilistic approach for demand-aware
  ride-sharing optimization,'' in \emph{{ACM} International Symposium on Mobile
  Ad Hoc Networking and Computing}, 2019.

\bibitem{Fang2017}
Z.~Fang, L.~Huang, and A.~Wierman, ``Prices and subsidies in the sharing
  economy,'' in \emph{International Conference on World Wide Web}, 2017.

\bibitem{fang2020loyalty}
Z.~Fang, L.~Huang, and A.~Wierman, ``Loyalty programs in the sharing economy:
  Optimality and competition,'' \emph{Performance Evaluation}, vol. 143, p.
  102105, 2020.

\bibitem{Ouyang2019}
Y.~Ouyang, B.~Guo, X.~Lu, Q.~Han, T.~Guo, and Z.~Yu, ``{CompetitiveBike}:
  Competitive analysis and popularity prediction of bike-sharing apps using
  multi-source data,'' \emph{{IEEE} Transactions on Mobile Computing}, vol.~18,
  no.~8, pp. 1760--1773, aug 2019.

\bibitem{Peng2018}
Z.~Peng, W.~Shan, P.~Jia, B.~Yu, Y.~Jiang, and B.~Yao, ``Stable ride-sharing
  matching for the commuters with payment design,'' \emph{Transportation},
  2018.

\bibitem{Fraiberger2015}
S.~P. Fraiberger and A.~Sundararajan, ``Peer-to-peer rental markets in the
  sharing economy,'' \emph{{SSRN} Electronic Journal}, 2015.

\bibitem{xu2021impact}
X.~Xu, S.~Zeng, and Y.~He, ``The impact of information disclosure on consumer
  purchase behavior on sharing economy platform airbnb,'' \emph{International
  Journal of Production Economics}, vol. 231, p. 107846, 2021.

\bibitem{D3}
Y.~Lan, J.~Peng, F.~Wang, and C.~Gao, ``Quality disclosure with information
  value under competition,'' \emph{International Journal of Machine Learning
  and Cybernetics}, vol.~9, no.~9, pp. 1489--1503, 2018.

\bibitem{D4}
J.~Zhang and K.~J. Li, ``Quality disclosure under consumer loss aversion,''
  \emph{Management Science}, vol.~67, no.~8, pp. 5052--5069, 2021.

\bibitem{D5}
X.~Bian, J.~C. Contat, B.~D. Waller, and S.~A. Wentland, ``Why disclose less
  information? toward resolving a disclosure puzzle in the housing market,''
  \emph{The Journal of Real Estate Finance and Economics}, pp. 1--44, 2021.

\bibitem{D6}
Y.~Duan, X.~Ruan, and C.~Chen, ``Information-disclosing strategies of
  third-party sellers on retail platforms,'' \emph{Managerial and Decision
  Economics}, 2021.

\bibitem{D1}
L.~Zhao, C.~Wang, X.~Peng, B.~Liu, and D.~Ahlstrom, ``The pricing strategy of
  oligopolistic competition food firms with the asymmetric information and
  scientific uncertainty,'' \emph{Journal of Food Quality}, 2017.

\bibitem{D2}
L.~Shao, J.~K. Ryan, and D.~Sun, ``Responsible sourcing under asymmetric
  information: price signaling versus supplier disclosure,'' \emph{Decision
  Sciences}, vol.~51, no.~5, pp. 1082--1109, 2020.

\bibitem{BOARD2009}
O.~Board, ``{Competition} {And} {Disclosure},'' \emph{The Journal of Industrial
  Economics}, vol.~57, no.~1, pp. 197--213, 2009.

\bibitem{Cheong2004}
I.~Cheong and J.-Y. Kim, ``Costly information disclosure in oligopoly,''
  \emph{Journal of Industrial Economics}, vol.~52, no.~1, pp. 121--132, 2004.

\bibitem{Janssen2016}
M.~C. Janssen and M.~Teteryatnikova, ``Horizontal product differentiation:
  Disclosure and competition,'' \emph{The Journal of Industrial Economics},
  vol.~64, no.~4, pp. 589--620, 2016.

\bibitem{Stivers2004}
A.~E. Stivers, ``Unraveling of information: Competition and uncertainty,''
  \emph{Topics in Theoretical Economics}, vol.~4, no.~1, 2004.

\bibitem{Dedman2009}
E.~Dedman and C.~Lennox, ``Perceived competition, profitability and the
  withholding of information about sales and the cost of sales,'' \emph{Journal
  of Accounting and Economics}, vol.~48, no. 2-3, pp. 210--230, 2009.

\bibitem{policy}
``Airbnb's guest refund policy for homes.''
  \url{https://www.airbnb.com/help/article/544/what-is-airbnbs-guest-refund-policy-for-homes?locale=en&_set_bev_on_new_domain=1575857748_MGY1MzM2Zjg3MGZh},
  2019.

\bibitem{term}
``Taskrabbit terms of service.'' \url{https://www.taskrabbit.com/terms}, 2019.

\bibitem{Christenfeld1995}
N.~Christenfeld, ``Choices from identical options,'' \emph{Psychological
  Science}, vol.~6, no.~1, pp. 50--55, 1995.

\bibitem{TEIGEN1983}
K.~H. Teigen, ``Studies in subjective probability l: Prediction of random
  events,'' \emph{Scandinavian Journal of Psychology}, vol.~24, no.~1, pp.
  13--25, 1983.

\bibitem{Li2017}
M.~Li, N.~C. Petruzzi, and J.~Zhang, ``Overconfident competing newsvendors,''
  \emph{Management Science}, vol.~63, no.~8, pp. 2637--2646, 2017.

\bibitem{Ren2013}
Y.~Ren and R.~Croson, ``Overconfidence in newsvendor orders: An experimental
  study,'' \emph{Management Science}, vol.~59, no.~11, pp. 2502--2517, 2013.

\bibitem{Liang2017}
L.~J. Liang, H.~C. Choi, and M.~Joppe, ``Understanding repurchase intention of
  airbnb consumers: perceived authenticity, electronic word-of-mouth, and price
  sensitivity,'' \emph{Journal of Travel {\&} Tourism Marketing}, vol.~35,
  no.~1, pp. 73--89, 2017.

\bibitem{Tussyadiah2014}
I.~P. Tussyadiah, ``An exploratory study on drivers and deterrents of
  collaborative consumption in travel,'' in \emph{Information and Communication
  Technologies in Tourism 2015}, 2014, pp. 817--830.

\bibitem{So2018}
K.~K.~F. So, H.~Oh, and S.~Min, ``Motivations and constraints of airbnb
  consumers: Findings from a mixed-methods approach,'' \emph{Tourism
  Management}, vol.~67, pp. 224--236, 2018.

\bibitem{Verrecchia1983}
R.~E. Verrecchia, ``Discretionary disclosure,'' \emph{Journal of Accounting and
  Economics}, vol.~5, pp. 179--194, 1983.

\bibitem{Ghosh2015}
A.~Ghosh and A.~Roth, ``Selling privacy at auction,'' \emph{Games and Economic
  Behavior}, vol.~91, pp. 334--346, 2015.

\bibitem{Ghosh2014}
A.~Ghosh, K.~Ligett, A.~Roth, and G.~Schoenebeck, ``Buying private data without
  verification,'' in \emph{{ACM} conference on economics and computation},
  2014.

\bibitem{zhou2008preserving}
B.~Zhou and J.~Pei, ``Preserving privacy in social networks against
  neighborhood attacks,'' in \emph{IEEE 24th International Conference on Data
  Engineering}, 2008.

\bibitem{hampton2007neighborhoods}
K.~N. Hampton, ``Neighborhoods in the network society the e-neighbors study,''
  \emph{Information, Communication \& Society}, vol.~10, no.~5, pp. 714--748,
  2007.

\end{thebibliography}

\vfill

\end{document}